\documentclass[12pt]{article}
\usepackage{amsmath,amssymb}
\usepackage{graphicx}
\usepackage{subfigure}
\usepackage{color}

\newcommand{\la}{\langle}
\newcommand{\ra}{\rangle}
\newcommand{\ol}{\overline}
\newcommand{\mq}{m_{\tilde{q}}}
\newcommand{\mg}{m_{\tilde{g}}}

\newcommand{\Slash}[1]{{\ooalign{\hfil/\hfil\crcr$#1$}}} 
\newcommand{\lpartial}{\hspace{0.1em}\raisebox{1ex}{$\leftarrow$}\hspace{-0.85em}\raisebox{-.6ex}{$\partial$}\hspace{0.3em}}
\makeatletter

\@addtoreset{equation}{section}
\makeatother

\begin{document}\baselineskip=16pt
\begin{titlepage}
\begin{flushright}
{\small OU-HET 725/2011}\\
\end{flushright}
\vspace*{1.2cm}

\begin{center}

{\Large\bf 
QCD parity violation in a quarkonium via SUSY\\
} 
\lineskip .75em
\vskip 1.5cm

\normalsize
{\large Naoyuki Haba},
{\large Kunio Kaneta},
and 
{\large Tetsuya Onogi}

\vspace{1cm}

{\it Department of Physics, 
 Osaka University, Toyonaka, Osaka 560-0043, 
 Japan} \\

\vspace*{10mm}

{\bf Abstract}\\[5mm]
{\parbox{13cm}{\hspace{5mm}
%

The supersymmetric standard model
 undergoes parity violation in QCD through 
 chiral    
 quark-squark-gluino 
 interactions with 
 non-degenerate masses between 
 left-handed and right-handed 
 squarks. 
Since experiments have not shown any parity violation  
 in QCD yet,  
 a bound for the mass degeneracy between 
 left-handed and right-handed squarks   
 should exist. 
In this paper we try to obtain this 
 bound for each squark. 
Firstly, we investigate a non-degeneracy bound between 
 $m_{\tilde{c}_L}$ and  $m_{\tilde{c}_R}$ from 
 experimental data of charmonium decay.  
 Secondly, we estimate the non-degeneracy bounds for
 $\tilde{u}$ and $\tilde{d}$ from nucleon-meson 
 scattering data, 
 and comment on other squarks. 
\newline

}}

\end{center}

\end{titlepage}

\section{Introduction}

The supersymmetric (SUSY) standard model (SM) 
 is one of the most promising candidates beyond 
 the SM due to gauge coupling unification, 
 the possible existence of dark matter, and so on.  
It is worth noting that 
 the SUSY SM 
 has chiral gauge interactions 
 in fermion-sfermion-gaugino vertexes,  
 for example, 
 left-handed fermions only couple to 
 left-handed sfermions with gauginos. 
The left-right index in sfermions is just a label 
 and has nothing to do with spin, however, 
 fermion-sfermion-gaugino interactions 
 are exactly chiral. 
Moreover, 
 the mass of the left-handed sfermion is not the same 
 as that of the right-handed sfermion in general,
 and  
 this non-degeneracy is also induced from 
 radiative corrections picking up weak interactions. 
Therefore, 
 due to this non-degeneracy, 
 SUSY gauge interactions cause parity violation 
 even in QCD! 
This is a remarkable feature of 
 the SUSY SM, and this effect is negligible 
 in a lot of 
 other candidates beyond the SM\cite{qcd-pv}. 
We take a setup of 
 $R$-parity conservation, 
 where sparticles propagate only inside loop diagrams   
 due to their heavy masses. 
Since any parity violation has not been discovered 
 in QCD,  
 non-degeneracy bounds 
 should be obtained between 
 the left- and right-handed squark masses. 
 
Does 
 the experimental fact of 
 conservation of parity in QCD 
 suggest a degeneracy 
 between left- and right-handed squarks? 
We must investigate this degeneracy, and 
 try to obtain non-degeneracy bounds from 
 current experimental data. 
As for the degeneracy of $\tilde{t}_L$ and $\tilde{t}_R$, 
 there have been some research on the process of 
 $t \bar{t}$ pair production in collider
 experiments \cite{qcd-pv,Berge:2007dz}. 
 Amplitudes of  $t \bar{t}$
 pair production were found their depend on 
 helicities,
 and 
 non-degeneracy between 
 $\tilde{t}_L$ and $\tilde{t}_R$ 
 causes the   
 asymmetry measurement of the cross section. 
The case of ${\mathcal O}(100)$ GeV (${\mathcal O}(1)$ TeV)
 masses of sparticles  
 was investigated in 
 Ref. \cite{Berge:2007dz} (Ref. \cite{qcd-pv}).

Here we should comment on 
 sparticle masses which are consistent 
 with collider 
 experiments. 
One case is that 
 all sparticles are 
 heavy of 
 ${\mathcal O}(1)$ TeV 
 as well as
 gluino mass $> 600$ GeV \cite{daCosta:2011qk},
 where all sparticles are too heavy 
 to be detected at detectors in 
 current experiments. 
Another case 
 is that 
 light sparticles exist which  
 are degenerate to 
 $30$ GeV 
 compared to other
 heavy sparticles. 
It is because 
 there are experimental cuts for $p_T$s multi-jets 
 with missing transverse momentum 
 in the SUSY search at the LHC (Tevatron), where
 an event selection for jets is
 $p_T>40$ GeV \cite{daCosta:2011qk} ($p_T>30$ GeV \cite{:2007ww}), 
 and 
 $p_T$ of jets are roughly estimated as the mass difference
 of gluino and squarks. 
Thus, 
 the degeneracy of sparticles within 30 GeV
 is consistent with collider experiments. 
We will consider both cases 
 in the following numerical analyses.

In this paper, 
 we try to obtain the 
 bound for left-right degeneracy of squark masses 
 other than the stop. 
At first,
 we investigate a non-degeneracy bound between 
 $m_{\tilde{c}_L}$ and  $m_{\tilde{c}_R}$ from 
 experimental data of charmonium decay.  
For this analys, 
 we use non-relativistic QCD (NRQCD) \cite{Bodwin:1994jh},  
 since 
 the charmonium is heavy. 
The $q\bar q$ bound state in NRQCD is considered in Section 2,
and some related calculations are shown in Appendix A.
We utilize dimension six operators to evaluate a non-degeneracy bound.
Dimension six operators are discussed in Appendix B.
In Section 3 and 4,
we calculate a formula for the decay width by using an effective action technique.
In Section 5, we estimate bounds for left-right non-degeneracy of $\tilde{c}$.
We also estimate the non-degeneracy bounds for
 $\tilde{u}$ and $\tilde{d}$ from nucleon-meson 
 scattering data, 
 and finally comment on bounds for 
 other squarks.


\section{Quarkonium in NRQCD}

Let us 
 consider a quarkonium of 
 $q\ol{q}$ bound
 state in the NRQCD framework 
 by introducing a bilocal field. 
It is applicable for heavy quarks, and 
 a related work has been shown in,
 for example, 
 Refs. \cite{Kleinert:1976xz,Kugo:1978ct,Morozumi:1986zx,Hisano:2004ds,Brambilla:2004wf,Brambilla:2004jw,Brambilla:2010cs}. 

A two-body effective action in NRQCD is given by 
\begin{align}
S_{\rm eff}&=
\int_X \int_{\vec{r}}
\phi^{\mu\dagger}_X(\vec{r})
\left[
i\partial^0_X-\frac{\nabla^2_X}{4m}+H(r)
\right]
\phi_{\mu X}(\vec{r}),\label{Seff}
\end{align}
where $H(r)$ is defined as $H(r)\equiv -\nabla_r^2/m-V(r)$.
A detail derivation of Eq. (\ref{Seff}) is given in Appendix A.
Now we estimate the spectra of bound states $\phi^\mu_X(\vec r)$. 
%
 $\phi^\mu_X(\vec{r})$ can be expanded 
 by a complete set of 
 $\psi_n(\vec{r})$ as 
\begin{align}
\phi^\mu_X(\vec{r})
=\sum_n a^\mu_n(X)\psi_n(\vec{r})
=\sum_n \int\frac{d^3P}{(2\pi)^3}a^\mu_n(\vec{P})\psi_n(\vec{r})e^{-iP\cdot X}
,
\label{expansion}
\end{align}
where 
 $a^\mu_n(X)$ is a plane wave, and 
 $\psi_n(\vec{r})$ is a possible bound state 
 which this system can take. 
An eigenstate of $H(\vec{r})$, 
 which satisfies    
\begin{align}
\hat{H}(\hat{\vec{r}})\psi_n(\vec{r})
&=
E_n\psi_n(\vec{r}), 
\label{schrodingereq}
\end{align}
is a quarkonium, 
 and 
 $E_n$ denotes a binding energy of it. 
Orthogonality and completness suggest 
\begin{align}
\int d^3r \psi^{\dagger}_n(\vec{r})\psi_m(\vec{r})
=\delta_{nm}, \;\;\;
\sum_n\psi_n(\vec{r})\psi^{\dagger}_n(\vec{s})
=\delta(\vec{r}-\vec{s}). \label{closure}
\end{align}
A hadron wave function is factorized by 
 $a^\mu_n(X)$, which only depends on  
 center of mass coordinate. 
Here 
 $\mu$ represents spin singlet (triplet) state of meson 
 when $\mu=0$ ($\mu=i$).  
Note that
 a hadron labeled by $n$ is created by 
 $a^{\mu\dagger}_n(X)$ as 
 $a^{\mu\dagger}_n(X)|0\rangle=|n\rangle$.

Here let us apply 
 this formalism to 
 a charmonium, for example. 
We denote 
 $n=\eta_c, h_c, J/\psi,$ $\chi_c, \cdots$,  
 then 
 a spin singlet state 
 $\phi^0_X(\vec{r})$ and 
 a spin triplet state
 $\phi^i_X(\vec{r})$ are 
 represented by 
\begin{align}
\phi^0_X(\vec{r})
&
=a^0_{\eta_c}(X)\psi_{\eta_c}(\vec{r})+a^0_{h_c}(X)\psi_{h_c}(\vec{r})+\cdots,\\
\phi^i_X(\vec{r})
&
=a^i_{J/\psi}(X) \psi_{J/\psi} (\vec{r})+a^i_{\chi_{cJ}}(X)\psi_{\chi_{cJ}}(\vec{r})+\cdots,
\end{align}
respectively. 
We now obtain 
 the effective action of charmonium  
 in the SM QCD, where parity is conserved.


\section{Direct parity violation}

In the SUSY SM, 
 parity can be violated in quarkonium 
 through the 
 non-degeneracy of left-right squark masses. 
As we have shown 
 in Appendix B, 
 there are three parity-violating operators, 
 ${\cal O}_{4F}^{(1)}$,  ${\cal O}_{4F}^{(8)}$, and 
 ${\cal O}_{qqG}$.   
At a direct decay vertex of quarkonium, 
 ${\cal O}_{4F}^{(1)}$ 
 gives the leading order of parity violation, 
 and we call this 
 process 
 ``direct parity violation''.  
%
The explicit form of 
 the direct parity violating operator 
 is given by  
\begin{align}
{\cal O}^{\rm p.v}_{4F}
&
=(A_{uc}+B_{cu})\delta^4(x-y)[\ol{u}(x)\gamma_\mu
 u(y)][\ol{c}(x)\gamma^\mu\gamma^5c(y)] ,
\label{pv4f}
\end{align}
where $A_{uc}$ and $B_{cu}$ are 
\begin{align}
A_{uc}
&\equiv
\frac{12g_s^4}{192\pi^2}\frac{1}{4}
(-C^{(\tilde{u},\tilde{c})}_{LL}+C^{(\tilde{u},\tilde{c})}_{RR}+C^{(\tilde{u},\tilde{c})}_{LR}-C^{(\tilde{u},\tilde{c})}_{RL}),\\
B_{cu}
&\equiv
\frac{12g_s^4}{192\pi^2}\frac{1}{4}
(-C^{(\tilde{c},\tilde{u})}_{LL}+C^{(\tilde{c},\tilde{u})}_{RR}-C^{(\tilde{c},\tilde{u})}_{LR}+C^{(\tilde{c},\tilde{u})}_{RL}),
\end{align}
respectively. 
We estimate $u$-quark contribution at first, and later 
 include $d$-quark contribution. 
Note that squark flavor is labeled by  
 $C^{(\tilde{q},\tilde{q}')}_{ij}$ $(i, j=L, R)$, 
 and has squark mass dependence through  
 $f_1(\mq,m_{\tilde{q}'})$ and $f_2(\mq,m_{\tilde{q}'})$. 
For example, 
 $C^{(\tilde{u},\tilde{c})}_{LL}$ is denoted as 
\begin{align}
C^{(\tilde{u},\tilde{c})}_{LL}
&
=\frac{2}{9}[f_1(m_{\tilde{u}_L},m_{\tilde{c}_L})+f_2(m_{\tilde{u}_L},m_{\tilde{c}_L})], 
\end{align}
and other $C$-factors 
 are similarly obtained by using Eqs.(\ref{22})$\sim$(\ref{24}).


As for 
 ${\cal O}_{qqG}$ and ${\cal O}_{4F}^{(8)}$, 
 they do not induce the leading order contributions, 
 because they 
 must emit a gluon in the decay vertex.   
We can neglect gluon exchange 
 between in-going and out-going states 
 at the decay instant in the NRQCD,   
 since 
 non-relativistic bound states 
 are hadronized by 
 space-like gluon exchanges. 
Therefore, 
 we can neglect 
 the contributions from 
 ${\cal O}_{qqG}$ and ${\cal O}_{4F}^{(8)}$, 
 and 
 factorize this decay process 
 by a vacuum insertion as in Fig. 1. 
 \begin{figure}[htbp]
 \begin{center}
 \includegraphics[keepaspectratio, scale=0.3]{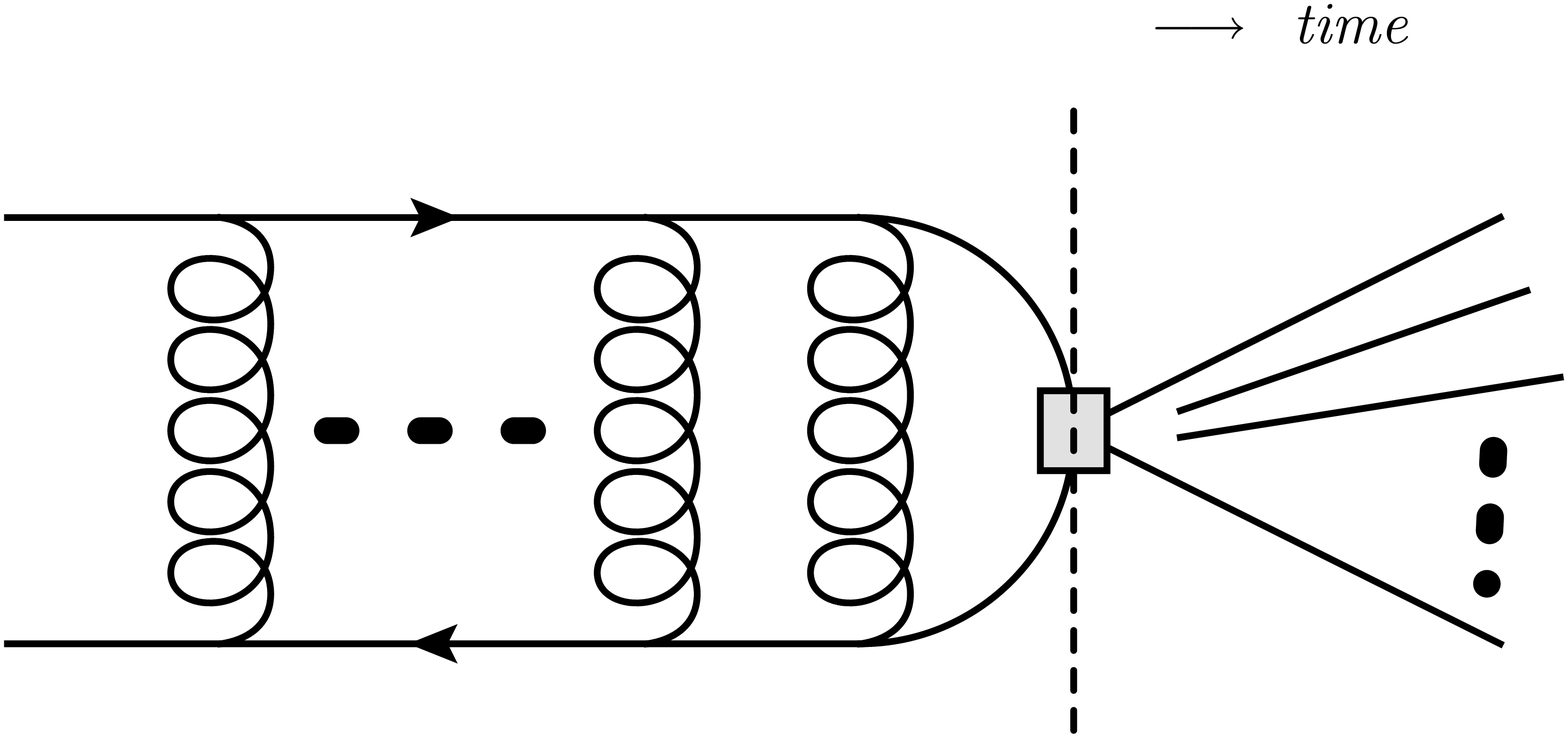}
 \label{direct}
 \caption{\footnotesize Factorization by a vacuum insertion 
 in a direct parity violation process. A box stands for 
 ${\cal O}^{\rm p.v}_{4F}$ in Eq.(\ref{pv4f}). }
 \end{center}
 \end{figure}
 
We focus on a charmonium, $\eta_c$, 
 which 
 is $0^{-+}$ under 
 $J^{PC}$, and 
 has mass of 2980 MeV. 
Notice that 
 ${\cal O}^{\rm p.v}_{4F}$ is 
 a contact interaction, where
 the decay constant is a value of wave function 
 at an origin due to $\delta$-function and  
 a decay through the contact interaction 
 is only 
 possible with the S-state (angular momentum $L=0$). 
Thus, 
 reminding $\pi$ is $0^{-+}$, 
 $\eta_c$ can not decay to $\pi\pi$ until 
 it pick up parity violation, 
 since  $\pi(p), \pi(-p)$ system\footnote{
 It has $P=(-1)^L$ and $C=(-1)^{S+L}$.}
 of S-state 
 is $0^{++}$.   
Note that there exits weak interaction, however,
 it also breaks $C$. 
Anyhow, as in Fig. 1, 
 the direct parity violation through the SUSY effects, i.e., 
 a two-body decay process, $\eta_c\rightarrow \pi\pi$,
 should be
 factorized as 
 $\langle \pi\pi |{\cal O}^{\rm p.v}_{4F}|\eta_c\rangle\sim\langle
 \pi\pi |\ol{q}\gamma^\mu q|0\rangle\langle
 0|\ol{q}\gamma_\mu\gamma^5q|\eta_c\rangle$. 
Here $\langle \pi\pi |\ol{q}\gamma^\mu q|0\rangle$
 is a pion form factor, 
 and we can estimate 
 $\langle 0|\ol{q}\gamma_\mu\gamma^5q|\eta_c\rangle$ 
 by use of NRQCD.  
Actually, 
 by regarding 
 $\ol{q}\gamma^0\gamma^5q \sim
 -\frac{1}{2}\chi^\dagger\varphi+{\rm h.c.}$ 
 in a non-relativistic picture, 
 the S-matrix element of  
 $\eta_c\rightarrow\pi\pi$ 
 is given by 
\begin{align}
\langle \pi\pi |{\cal O}^{\rm p.v}_{4F}|\eta_c\rangle
&\sim
-\frac{1}{2}(A_{uc}+B_{cu})\delta^4(x-y)
\langle \pi\pi |u^\dagger(x) u(y)|0\rangle
\langle0|\chi^\dagger(x)\varphi(y)|\eta_c\rangle .
\end{align}
Here 
 $\langle \pi\pi |u^\dagger(x) u(y)|0\rangle\sim F^s(k)$ 
 is a scalar form factor of pion, 
 which has non-trivial energy dependence.

In general, 
 when a bound state $|n\ra (\equiv a^{\nu\dagger}_n(P)|0\ra)$
 decays through 
 a bilocal operator
 ${\cal O}^{\nu\lambda\cdots}_X(\vec{r}) =
 \phi^{\mu}_X(\vec{r}) \phi^{\lambda}_X(\vec{r}) \cdots$, 
 its matrix element is given by 
\begin{align}
\la 0|T[{\cal O}^{\nu\lambda\cdots}_X(\vec{r})]|n\ra
&
=i\int d^4Y\int d^3sF^n_P(Y;s)
\left(
i\partial_{Y^0}-\frac{\nabla^2_Y}{4m}-\hat{H}(s)
\right)
\la 0|T[{\cal
 O}^{\nu\lambda\cdots}_X(\vec{r})\phi^{\mu\dagger}_Y(s)]|0\ra ,\nonumber
\label{trampx}
\end{align}
where 
$F^n_P(Y;s)\equiv \psi_n(\vec{s})e^{-iP\cdot Y}$,   
and it satisfies 
$\int d^3Xd^3r\phi^{\mu\dagger}_X(r)F^n_P(X;r)
=a^\dagger_n(\vec{P})$ 
{}from orthogonality and completeness. 
Thus, 
 the transition amplitude in Eq.(\ref{trampx}) 
 is given by 
 $\psi_n (\vec{r}) e^{-iPX}$ with 
 ${\cal O}^{\nu\lambda\cdots}_X(\vec{r})=\phi^\nu_X(\vec{r})$.

Let us go back to a charmonium, and 
 take $q$ as $c$-quark in Eq.(\ref{FWT}).   
Since 
 a heavy quark is non-relativistically expanded as Eq.(\ref{FWT}), 
 the 4-Fermi operator can be also expanded similarly.  
In the leading order, 
 components of 
 $\chi^\dagger \varphi$ and $\varphi^\dagger \chi$ 
 in the bilocal field,  
 are only 
 creating and annihilating 
 operators of charmonium.  
Thus, 
 $\phi^n_X(\vec{r})$ corresponds to 
 $\chi^\dagger(x)\varphi(y)$, 
 and we name a label $n=0$ 
 $\eta_c$ for the charmonium,   
 which suggests 
\begin{align}
\la 0|\phi^0_X(\vec{r})|\eta_c\ra
&
=\psi_{\eta_c}(\vec{r})e^{-iP\cdot X}.
\end{align}
Remind that 
 ${\cal O}^{\rm p.v}_{4F}$ is 
 a contact interaction, 
 and we can use 
 $m_{\eta_c}$ for an energy of the pion form factor 
 due to a momentum conservation.  
Then, 
 we  
 obtain 
\begin{align}
\la \pi\pi |{\cal O}^{\rm p.v}_{4F}|\eta_c\ra
&\sim
-\frac{1}{2}(A_{uc}+B_{cu})F^s(m_{\eta_c})\psi_{\eta_c}(0) .
\end{align}
There is 
 a $d$-quark contribution as well as
 $u$-quark ones, so that 
 the effective 4-Fermi operator ${\cal O}^{\rm p.v}_{4F}$ 
 becomes 
 a linear combination of $u$ and $d$.
Therefore, 
 $\Gamma(\eta_c\rightarrow \pi\pi)$
  is estimated as  
\begin{align}
\Gamma(\eta_c\rightarrow \pi\pi)
&\sim
\mid A_{uc}+A_{dc}+B_{cu}+B_{cd}\mid^2
 \frac{|F^s(m_{\eta_c})|^2|\psi_{\eta_c}(0)|^2}{16m^2_{\eta_c}}. 
\end{align}
Since $\eta_c$ is an S-state,
the decay width
 depends only on 
the  wave function at the origin .
This is a characteristic feature in 
 the direct parity violating process 
 in the SUSY SM.


\section{Indirect parity violation}

The QCD dimension six operators from the SUSY SM 
 can have 
 the parity violating 
 effects, and  
 actually, they can also 
 contribute 
 organization of quarkoniums themselves. 
We call this effect ``indirect parity violation'', 
 and we investigate it in this section.  
For this indirect parity violation, 
 all 
 ${\cal O}_{qqG},{\cal O}^{(1)}_{4F},$ and ${\cal O}^{(8)}_{4F}$
 contribute as 
 in Fig. 2. 

 \begin{figure}[htbp]
 \begin{center}
 \includegraphics[keepaspectratio, scale=0.4]{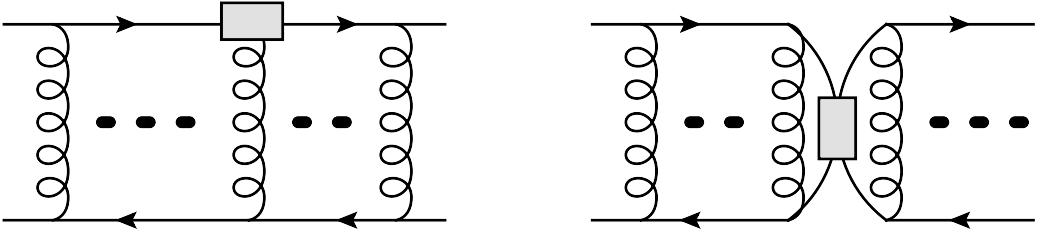}
 \label{indirect}
 \caption{\footnotesize Diagrams which contribute indirect parity violation through 
 dimension six operators (which are shown a box at a vertex). 
(Left): a contribution from ${\cal O}_{qqG}$, 
(Right): a contribution from
 ${\cal O}^{(1)}_{4F}$ or ${\cal O}^{(8)}_{4F}$. }
 \end{center}
 \end{figure}

The indirect parity violation induces a mixing between 
 an even-parity state and an odd-parity state as well as  
 a S-state and a P-state in a quarkonium. 
As 
 the parity violating term is written by 
 $\delta V^{\mu\nu}(r)$ in the potential, 
 the effective action in Eq.(\ref{NRQCD-Seff}) 
 includes indirect parity violation 
 by rewriting 
 $V(r)g^{\mu\nu}\rightarrow V(r)g^{\mu\nu}+\delta V^{\mu\nu}(r)$.   
Here 
 $\delta V^{\mu\nu}(r)$ is a matrix 
 in a basis of S- and P-states, 
 which has off-diagonal elements 
 of hadron state labeled by $n$ (and $\mu$). 
Now let us calculate the mixing between 
 asymptotic states in the SUSY SM 
 by using the basis of the SM QCD. 
Since the potential only depends on relative coordinate, 
 the wave function can be expanded by 
 $\Psi_n(\vec{r})$ in the SUSY SM 
 as 
\begin{align}
\phi^\mu_X(\vec{r})
&
=\sum_n A^{\mu}_n(X)\Psi_n(\vec{r}), \label{expansion}
\end{align}
where $\Psi_n(\vec{r})$ satisfies eignvalue equations,
\begin{align}
[H^{\rm QCD}(\vec{r})+\delta V(\vec{r})]\Psi_n(\vec{r})
&
=E^{\rm full}_n\Psi_n(\vec{r}), 
\end{align}
for $E^{\rm full}_n\neq  E_n$. 
Note that 
 $n$ is the label of the hadron, 
 which contains an information of spin
 ($\mu=0$: singlet, $\mu=i$: triplet).  
This $\Psi_n(\vec{r})$ must be 
 $\Psi_n(\vec{r})\rightarrow\psi_n(\vec{r})$ as 
 $\delta V\rightarrow 0$, so that it is given by 
 \begin{align}
\Psi_n(\vec{r})
&
=\psi_n(\vec{r})
+\sum_{k\neq n}\frac{V_{nk}}{E_n-E_k}\psi_k(\vec{r}), 
\end{align}
up to
 the first order of perturbation. 
Note that 
 $\Psi_n(\vec{r})$ must satisfy 
\begin{align}
\int d^3r\Psi^{\dagger}_n(\vec{r})\Psi_m(\vec{r})
&
=\delta_{mn}
\end{align}
for the zeroth order of perturbation. 
$V_{nk}$ is defined by 
\begin{align}
V_{nk}
&
\equiv
\int d^3s\psi^{\dagger}_k(\vec{s})[\delta V(\vec{s})]\psi_n(\vec{s}). 
\label{Vnk}
\end{align}
%
The classical complete set $\Psi_n(\vec{r})$ 
 should be
 written by 
 the QCD complete set $\psi_n(\vec{r})$,  
 while 
 a hadron creation operator is given by
 $A^{\mu\dagger}_n(X)$, 
 so that 
 $A^\mu_n(X)$ corresponds to $a^\mu_n(X)$. 
$\Psi_n(\vec{r})$ and $\psi_n(\vec{r})$ 
 are different complete bases as 
\begin{align}
\phi^\mu_X(\vec{r})
=\sum_n a^\mu_n(X)\psi_n(\vec{r}) 
=\sum_n A^\mu_n(X)\Psi_n(\vec{r}) . 
\end{align}
Thus, 
 by use of orthogonalization of $\Psi_n(\vec{r})$, 
 we obtain 
\begin{align}
A^{\mu\dagger}_n(X)
&
=a^{\mu\dagger}_n(X)+\sum_{k\neq
 n}\frac{(V_{nk}a^{\dagger}_{k}(X))^\mu}{E_n-E_k}\ . 
\label{mixing}
\end{align}

Let us consider a charmonium system. 
Equation (\ref{mixing}) means 
 an observed  
 $\eta_c$ is almost represented by a 
 mixing state of $\eta_c$ and
 $\chi_{c0}$ as 
\begin{align}
|\eta_c\ra_{\rm obs.}
&
=|\eta_c\ra+\frac{V_{\eta_c,
 \chi_{c0}}}{E_{\eta_c}-E_{\chi_{c0}}}|\chi_{c0}\ra .
\label{mixing1}
\end{align}
$\chi_{c0}$ is $0^{++}$ with mass of 3415 MeV, and  
 a decay of $\chi_{c0}$ to $\pi \pi$ is possible 
 (see, Eq.(\ref{chipipi})) 
 when $\pi$-$\pi$ system has angular momentum, $L=1$. 
We estimate parity violating potential 
 induced from the SUSY SM. 
As for 
 ${\cal O}^{(1)}_{4F}$ in Fig. 2, 
 its coefficient 
 only depends on 
 $m_{\tilde{c}_L}$ and $m_{\tilde{c}_R}$,
 since 
 the bound state is charmonium. 
The parity violating terms in 
 ${\cal O}^{(1)}_{4F}$ 
 are given by 
\begin{align}
{\cal O}^{(1)}_{4F}
&\supset
\frac{12g_s^4}{192\pi^2}\frac{1}{2}(-C^{(\tilde{c},\tilde{c})}_{LL}+C^{(\tilde{c},\tilde{c})}_{RR})\delta^4(x-y)
[\ol{c}(x)\gamma^\mu c(x)]
[\ol{c}(y)\gamma_\mu\gamma^5 c(y)] ,
\end{align}
where we use spin relation,  
$
\delta_{\alpha\beta}\delta_{\gamma\lambda}
=\frac{1}{2}\delta_{\alpha\lambda}\delta_{\gamma\beta}
+\frac{1}{2}\sigma^a_{\alpha\lambda}\sigma^a_{\gamma\beta}$, and 
$\sigma^a\sigma^b
=\delta^{ab}+i\epsilon^{abc}\sigma^c$.  
A 
 color factor 
 is rewritten as 
$
\frac{1}{2}\delta_{ij}\delta_{kl}
=\frac{1}{2N_C}\delta_{il}\delta_{kj}+T^A_{il}T^A_{kj}$
 for an exchange of spin. 
We must be careful for exchanges of spin and 
 coordinate, 
 where only spin-singlet changes its sign (Table.\ref{SP}). 
\begin{table}[htdp]
\begin{center}
\begin{tabular}{|c|c|c|}
\hline
& exchange of spin ($\varphi \leftrightarrow \chi$)
 & exchange of coordinate ($x \leftrightarrow y$)\\
\hline
spin singlet $\phi^0(x,y)$&asym.&sym.\\
spin triplet $\phi^i(x,y)$&sym.&asym.\\
\hline
\end{tabular}
\end{center}
\caption{exchanges of spin or coordinate}
\label{SP}
\end{table}%
After careful calculations, 
 ${\cal O}^{(1)}_{4F}$ is 
 given by 
\begin{align}
{\cal O}^{(1)}_{4F}
&\rightarrow
\frac{12g_s^4}{192\pi^2}
\frac{1}{2}(-C^{(\tilde{c},\tilde{c})}_{LL}+C^{(\tilde{c},\tilde{c})}_{RR})
\left(
\frac{i}{4m_cN_C}
\right)\nonumber\\
&\times
\left(
\begin{array}{c}
\phi^0\\
\phi^i
\end{array}
\right)^\dagger_{x,y}
\left(
\begin{array}{cc}
0&4{\cal V}(r)\partial^j_r\\
4\lpartial^i_r {\cal V}(r)&4i\epsilon^{ijk}
\lpartial^k_r{\cal V}(r)
\end{array}
\right)
\left(
\begin{array}{c}
\phi^0\\
\phi^j
\end{array}
\right)_{x,y},
\end{align}
where
 ${\cal V}\equiv \delta^4(x-y)$ and   
 $\phi^i\lpartial^i_r\equiv -\partial^i_r\phi^i$.
As for ${\cal O}^{(8)}_{4F}$, 
 we can use the calculation result of ${\cal O}^{(1)}_{4F}$, 
 since spin structure is the same. 
The different point is just color factor, and 
 by using 
$T^A_{ij} T^A_{kl}
=\frac{C_F}{2N_C}\delta_{il}\delta_{kj}-\frac{1}{N_C}T^A_{il}T^A_{kj}$, 
 we show 
 color octet part is 
 $C_F$ ($=(N_C^2-1)/(2N_C)$)  
 times larger    
 than ${\cal O}^{(1)}_{4F}$. 
Then, 
 non-relativistic potential from  
 ${\cal O}^{(1)}_{4F}$ and ${\cal O}^{(8)}_{4F}$ 
 with parity violation 
 is totally given by 
\begin{align}
&\delta V^{4F}_{\mu\nu}(r)\nonumber\\
&
=
\frac{12g_s^4}{192\pi^2}\frac{i}{8m_cN_C}
\left[
(-C^{(\tilde{c},\tilde{c})}_{LL}+C^{(\tilde{c},\tilde{c})}_{RR})
+C_F(-D^{(\tilde{c},\tilde{c})}_{LL}+D^{(\tilde{c},\tilde{c})}_{RR})
\right]
\left(
\begin{array}{cc}
0&4{\cal V}(r)\partial^j_r\\
4\lpartial^i_r {\cal V}(r)&4i\epsilon^{ijk}
\lpartial^k_r{\cal V}(r)
\end{array}
\right). 
\label{V-4F}
\end{align}

For 
 a non-relativistic potential from 
 ${\cal O}_{qqG}$ in Eq.(\ref{SUSY-qqG}),  
 we estimate leading part. 
Since ${\cal O}_{qqG}$  
 is not the contact interaction as ${\cal O}^{(1)}_{4F}$,
 its parity violation effects should be added 
 to the gluon potential. 
The bilocal operator after integrating out 
 gluon is given by 
\begin{align}
{\cal L}
&\sim
\frac{g_s^2}{96\pi^2C_F}
\left[
\ol{q}(x)T^AE^0_{L,R}P_{L,R}q(x)
\right]
V(r)
\left[
\ol{q}(y)T^A\gamma^0q(y)
\right], 
\end{align}
where 
 $E^0_{L,R}$ has eight terms in total, which are 
 categorized as 
\begin{align}
{\rm (i)}&\;\;\;
\pm e_1(\mq)\frac{g_s^2}{96\pi^2C_F}
\left[
(\ol{q}_xT^A\gamma_\mu\partial^\mu\partial^0q_x)
+(\partial^\mu\partial^0\ol{q}_xT^A\gamma_\mu q_x)
\right]
V(r)
\left[
\ol{q}(y)T^A\gamma^0q(y)
\right],\\
{\rm (ii)}&\;\;\;
\pm e_2(\mq)\frac{g_s^2}{96\pi^2C_F}
\left[
(\partial^\mu\ol{q}_xT^A\gamma_\mu\partial^0q_x)
+(\partial^0\ol{q}_xT^A\gamma_\mu\partial^\mu q_x)
\right]
V(r)
\left[
\ol{q}(y)T^A\gamma^0q(y)
\right],\\
{\rm (iii)}&\;\;\;
\pm\frac{g_s^2}{96\pi^2C_F}
\left[
 e_3(\mq)
 \left\{
(\ol{q}_xT^A\gamma^0\partial^2q_x)
+(\partial^2\ol{q}_xT^A\gamma^0q_x)
\right\}
+ e_4(\mq)
(\partial^\mu\ol{q}_xT^A\gamma^0\partial_\mu q_x)
\right]\nonumber\\
&\hspace{2cm}\times
V(r)
\left[
\ol{q}(y)T^A\gamma^0q(y)
\right],\\
{\rm (iv)}&\;\;\;
\pm (-e_5(\mq))\frac{g_s^2}{96\pi^2C_F}
i\epsilon^{\alpha\beta 0\nu}
\left[
\partial_\beta\ol{q}_xT^A\gamma_\nu\partial_\alpha q_x
\right]
V(r)
\left[
\ol{q}(y)T^A\gamma^0q(y)
\right].
\end{align}
Here, 
 sign 
 $+$ ($-$) means that quark chirality is R (L). 
In the non-relativistic 
 limit, (i) and (ii) vanish, since  
 components of $\mu=0$ and $\mu=i$ are cancelled 
 with each other.   
For this calculation, we have used a 
 NRQCD result, 
 $\partial^0q\sim {\cal O}((m_c^2v)^{3/2})$ 
 ($v$: $c$-quark velocity, $m_c$: $c$-quark mass).  
Actually, 
 (iii) 
 induces the 
 leading effects for the 
 potential. 
By 
 taking 
 leading order of $v$, 
 a power counting shows 
\begin{align}
\delta V^{qqG}_{\mu\nu}(r)
&
=[(e_4(m_{\tilde{q}_R})-e_4(m_{\tilde{q}_L}))-2(e_3(m_{\tilde{q}_R})-e_3(m_{\tilde{q}_L}))]\nonumber\\
&\times
\frac{g_s^2}{96\pi^2}
\left(
\frac{-im_q}{8N_C}
\right)
\left(
\begin{array}{cc}
0&V(r)\partial^j_r+\lpartial^j_rV(r)\\
V(r)\partial^i_r+\lpartial^i_rV(r)&
i\epsilon^{ijk}[V(r)\partial^k_r+\lpartial^k_rV(r)]
\end{array}
\right),
\end{align}
where we use color factor ($C_F/(2N_C)$) from 
 Fierz transformation. 
As for (iv), 
 $\alpha,\beta$ must be {\it space}-index, 
 so that the second derivative of {\it space}-index
 appears, which corresponds to D-state
 (or higher angular momentum states), so that it  
 does not contribute the mixing between S- and P-states. 
The (iv) does not contribute 
 the mixing between S- and P-states, too. 
Thus, 
 the leading order of
 parity violating potential, which 
 triggers the mixing
 between S- and P-states, is given by 
\begin{align}
\delta V^{\rm SUSY}_{\mu\nu}(r)
&
=\delta V^{4F}_{\mu\nu}(r)+\delta V^{qqG}_{\mu\nu}(r). 
\end{align}

Then, 
 we can calculate 
 $V_{\eta_c, \chi_{c0}}$ in a charmonium,
 and 
 a formula of decay width is given by  
\begin{align}
\Gamma(\eta_c\to \pi\pi)
&\sim
\left|
\frac{V_{\eta_c, \chi_{c0}}}{E_{\eta_c}-E_{\chi_{c0}}}
\right|^2
\Gamma(\chi_{c0}\to \pi\pi). 
\label{574}
\end{align}
A wave function of charmonium 
 is given by 
$\psi(\vec r) = R_n (r) Y_{lm}(\theta, \phi)$,
where $R_n (r)$ satisfies the Schr\"odinger equation (\ref{schrodingereq}) with Coulomb plus linear potential 
(Cornell potential),
\begin{eqnarray}
V(r) &=& -\frac{\kappa}{r} + \frac{r}{a^2}.
\label{cornell}
\end{eqnarray}
We take $\kappa = 0.52$ and $a = 2.34$ GeV$^{-1}$ for charmonium system \cite{Eichten:1978tg}.
Through the Schr\"odinger equation with this potential, 
we can obtain charmonium wave function numerically.
\section{Bounds for left-right non-degeneracy of squark masses}

We are in a stage to investigate 
 bounds for left-right non-degeneracy of squark masses. 
At first, 
 we investigate 
 bounds for $\tilde{c}$ by use of 
 calculation tools in this paper. 
Next, 
 we estimate bounds for $\tilde{u}$ and $\tilde{d}$ 
 by use of a similar technique in Ref.\cite{NuclParity}. 
And finally, 
 we comment on bounds for other 
 sfermions.

\subsection{Bound for $\tilde{c}$}

Let us investigate the left-right 
 non-degeneracy bound for the masses
 of $\tilde{c}_L$ and $\tilde{c}_R$ 
 by use of the calculation method shown above. 
For a charmonium, 
 we focus on 
 $\eta_c$, whose 
 decay has 
 upper bounds of 
 $P$ and $CP$ violations as \cite{Nakamura:2010zzi}  
\begin{eqnarray}
Br(\eta_c\to\pi^+\pi^-)<6.0 \times 10^{-4}, \;\;\;
Br(\eta_c\to\pi^0\pi^0)<4.0 \times 10^{-4}. 
\label{EXP}
\end{eqnarray}
Note again that 
 $\eta_c$ can not decay to $\pi\pi$ until 
 it picks up parity violation. 
On the other hand, 
 a branching ratio of 
 $\chi_{c0}\to\pi\pi$ is
\begin{eqnarray}
\label{chipipi}
&&Br(\chi_{c0}\to\pi\pi)=(8.4\pm 0.4)\times 10^{-3}. 
\end{eqnarray}

A branching ratio of $\eta_c\to\pi\pi$ in a  
 direct parity violation from Eq.(\ref{574}) is 
 given by 
 \begin{align}
Br(\eta_c\to\pi\pi)_{\rm dir.}
&=
\mid A_{uc}+A_{dc}+B_{cu}+B_{cd}\mid^2
 \frac{|F^s(m_{\eta_c})|^2|\psi_{\eta_c}(0)|^2}{16m^2_{\eta_c}\Gamma_{\eta_c}},
\end{align} 
where $\Gamma_{\eta_c}$ is the total decay width of $\eta_c$.  
Here we take a scalar form factor 
 of pion $F^s$ by an 
 input parameter as $F^s(m_{\eta_c}^2)=1,\ 0.1,\ 0.001$, 
 since 
 its theoretical estimation is difficult 
 above 1 GeV. 
On the other hand, 
 the indirect parity violation  
 in $\eta_c\to\pi\pi$  
 suggests 
\begin{align}
Br(\eta_c\to\pi\pi)_{\rm indir.}
&\sim
\left|
\frac{V^{\rm SUSY}_{\eta_c, \chi_{c0}}+V^{\rm EW}_{\eta_c, \chi_{c0}}}{E_{\eta_c}-E_{\chi_{c0}}}
\right|^2
Br(\chi_{c0}\to\pi\pi),
\end{align}
where $V^{\rm EW}_{\eta_c, \chi_{c0}}$ is the SM
 background induced from 
 a $Z$-boson exchange.  
It gives
 an additional effect 
 ${\cal V}(r)\equiv (\alpha/r)\exp (-m_Zr))$ 
 in Eq.(\ref{V-4F}), which 
 is shown as 
\begin{align}
\delta V^{\rm EW}_{\mu\nu}(r)
=
\frac{g^2}{\cos^2\theta_W}
\left(
\frac{1}{2}-\frac{2}{3}\sin^2\theta_W
\right)^2
\frac{iC_F}{8m_cN}
\left(
\begin{array}{cc}
0&4{\cal V}(r)\partial^j_r\\
4\lpartial^i_r {\cal V}(r)&4i\epsilon^{ijk}
\lpartial^k_r{\cal V}(r)
\end{array}
\right)   
\end{align}
in a basis of (S-state, P-state)   
 with 
 $N=2$ and $C_F=3/2$. 
Then, we can evaluate $V^{\rm EW}_{\eta_c, \chi_{c0}}$
with Eq.(\ref{Vnk}), 
and the branching ratio is given by
$Br(\eta_c\to \pi\pi)_{\rm SM}
\sim
\left|
\frac{V^{\rm EW}_{\eta_c, \chi_{c0}}}{E_{\eta_c}-E_{\chi_{c0}}}
\right|^2
Br(\chi_{c0}\to\pi\pi)\simeq 7.0\times 10^{-22}$.
%

In Figs. 3 and 4, 
 the branching ratios of $\eta_c\to\pi\pi$ from direct and indirect 
 parity violation effects are plotted, respectively,  
 where
 horizontal axis is a 
 magnitude of
 $(m_{\tilde{c}_L}^2-m_{\tilde{c}_R}^2)/m_{\tilde{g}}^2$.  
%
Note
 that the branching ratio from indirect parity violation 
 is larger than 
 that from direct parity violation. 
Unfortunately, 
 we can show that 
 the SUSY parity violating effect is smaller than 
 the experimental bound of Eq.(\ref{EXP}) 
 in the parameter region,  
 and  
 it is difficult to 
 obtain the non-degeneracy bound   
 between $m_{\tilde{c}_L}$ and $m_{\tilde{c}_R}$.
Figures 5 and 6 show 
 a case that $\tilde{g}$ and $\tilde{c}_R$ are degenerate
 around 650 GeV in mass. 
The magnitude of the horizontal axis is 
 varied from 
 $(m_{\tilde{c}_L}^2-m_{\tilde{c}_R}^2)/m_{\tilde{g}}^2 = 4.5$, 
 which is taken to be 
 consistent with LHC data.  
Notice that 
 the branching ratio becomes
 larger than that in Figs. 3 and 4, however, 
 the experimental bound is also much higher,
 and we can not obtain the bounds. 
%
\begin{figure}[htbp]
  \def\@captype{table}
  \begin{minipage}[c]{.48\textwidth}
 \includegraphics[width=80mm]{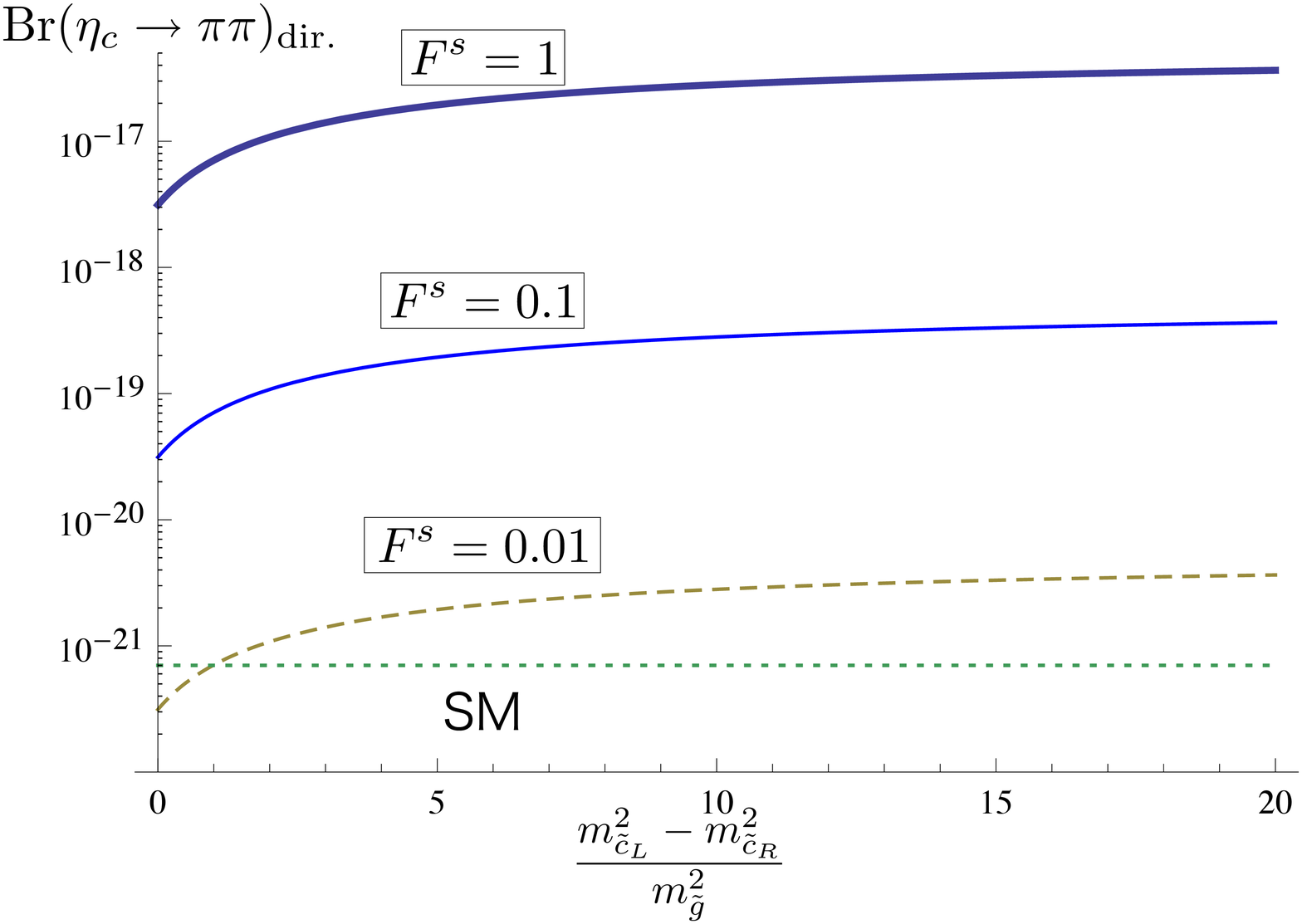}
\caption{\footnotesize Branching ratios of $\eta\to\pi\pi$ from direct parity
   violation with $\mg=1000\ {\rm GeV}$, $m_{\tilde{u}_R}=1600\ {\rm
   GeV}$, $m_{\tilde{u}_L}=2000\ {\rm GeV}$, $m_{\tilde{d}_R}=\ 1700{\rm
   GeV}$, $m_{\tilde{d}_L}=2100\ {\rm GeV}$, and $m_{\tilde{c}_R}=1800\
   {\rm GeV}$.} 
 \label{EtacDir}
\end{minipage}
  \hfill
  \begin{minipage}[c]{.48\textwidth}
\includegraphics[width=80mm]{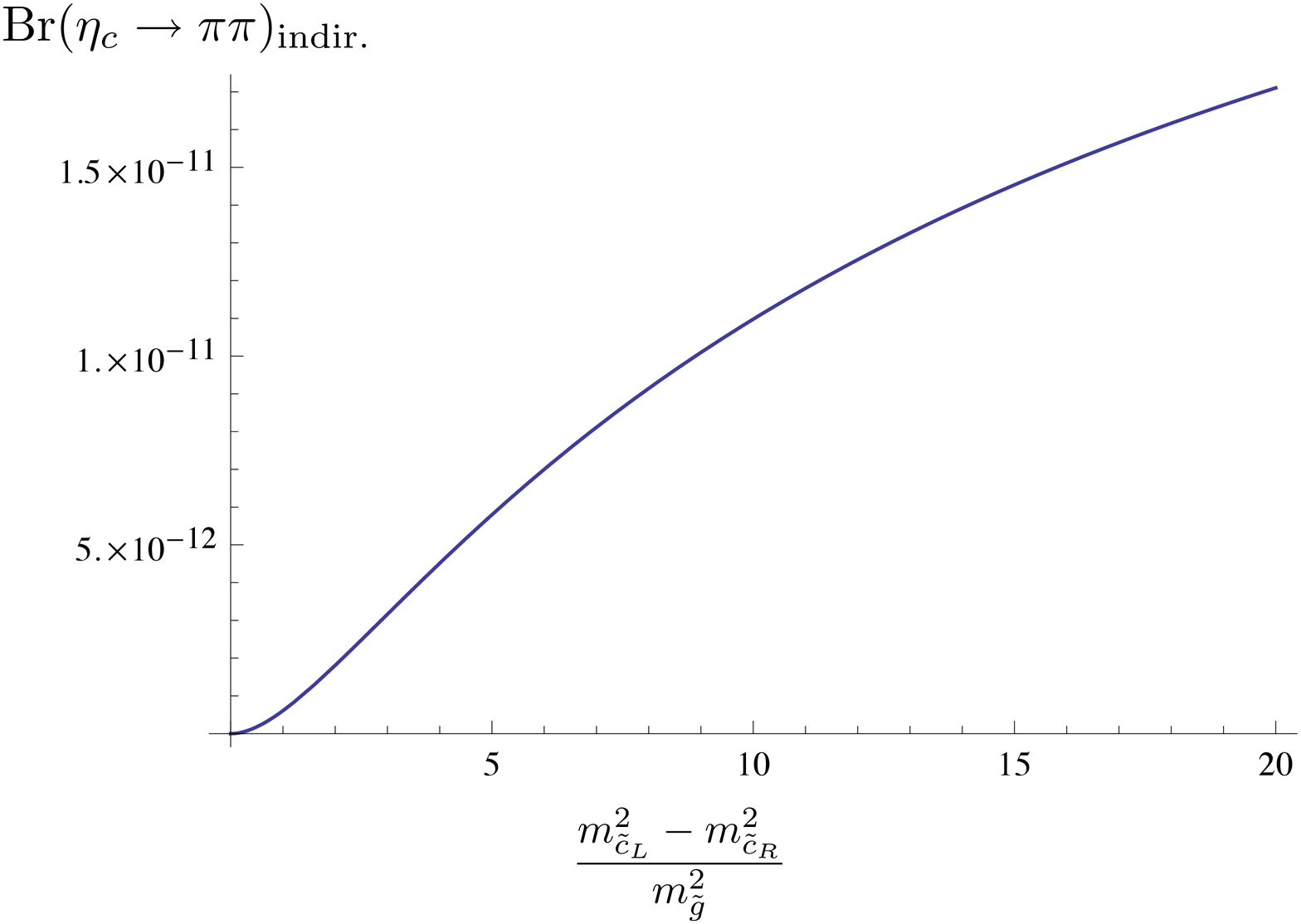}
\caption{\footnotesize Branching ratios of $\eta\to\pi\pi$ from indirect parity
   violation with
 $\mg=1000\ {\rm GeV}$, and $m_{\tilde{c}_R}=1800\ {\rm
   GeV}$.} 
 \label{EtacIndir}
\end{minipage}
 \end{figure}
\begin{figure}[htbp]
  \def\@captype{table}
  \begin{minipage}[c]{.48\textwidth}
\includegraphics[width=80mm]{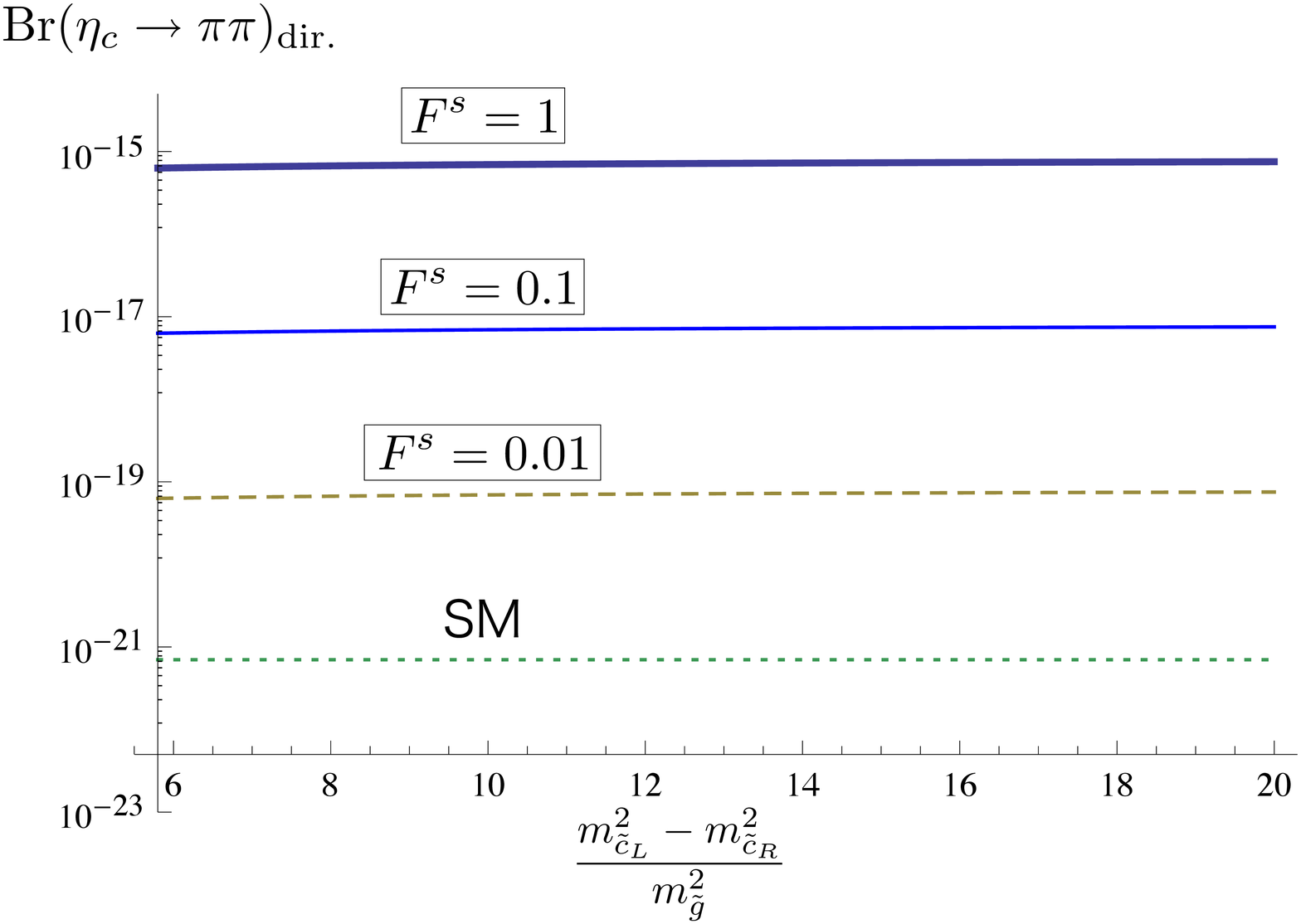}
\caption{\footnotesize Branching ratios of $\eta\to\pi\pi$ from indirect parity
   violation with
   $\mg=850\ {\rm GeV}$, $m_{\tilde{u}_R}=860\ {\rm GeV}$,
   $m_{\tilde{u}_L}=2000\ {\rm GeV}$, $m_{\tilde{d}_R}=\ 870{\rm GeV}$,
   $m_{\tilde{d}_L}=2100\ {\rm GeV}$, and $m_{\tilde{c}_R}=880\ {\rm
   GeV}$. } 
\label{EtacDir2}
\end{minipage}
  \hfill
  \begin{minipage}[c]{.48\textwidth}
\includegraphics[width=80mm]{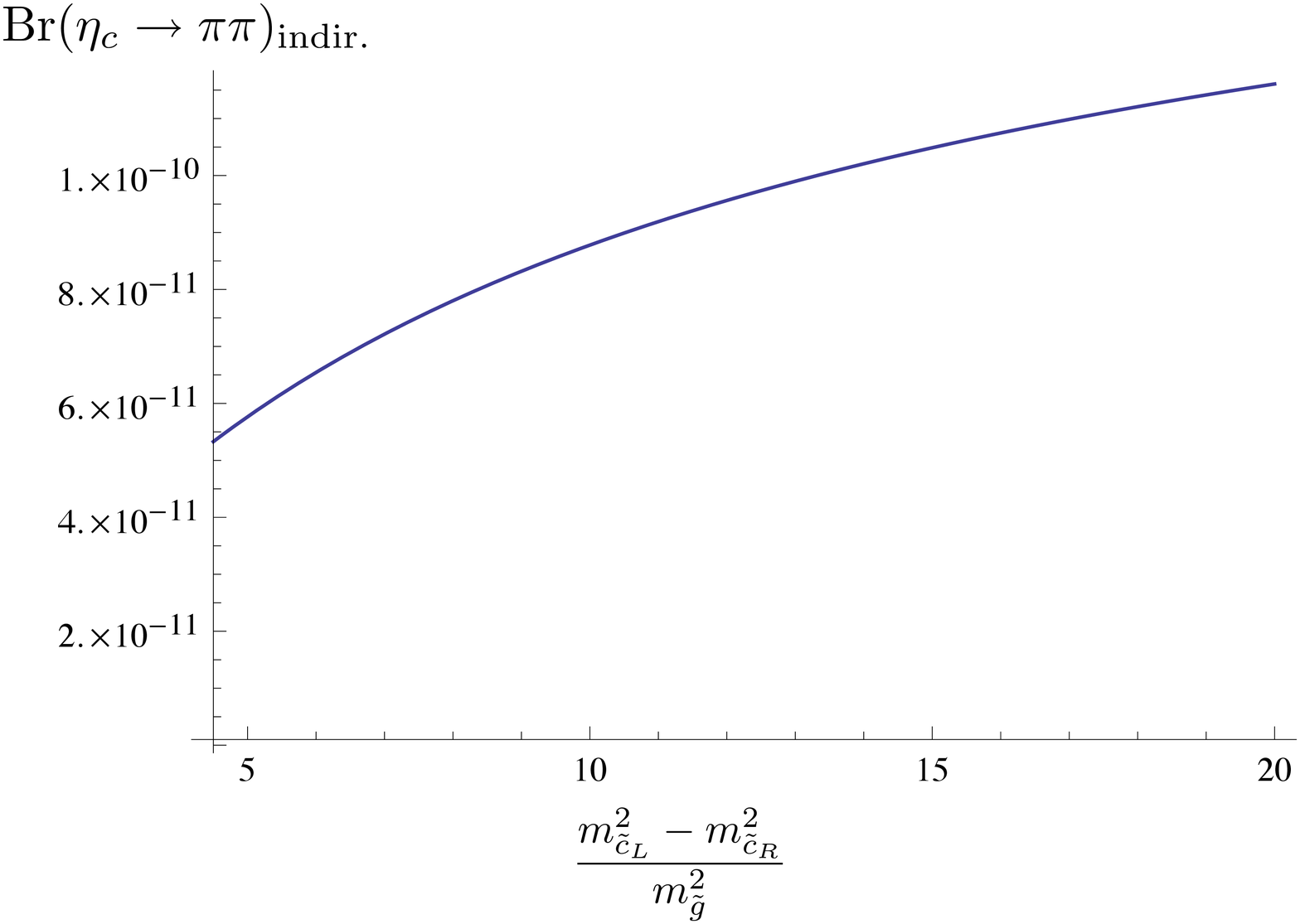}
\caption{\footnotesize Branching ratios of $\eta\to\pi\pi$ from indirect parity
   violation with
 $\mg=850\ {\rm GeV}$, and $m_{\tilde{c}_R}=880\ {\rm GeV}$.} 
\label{EtacIndir2}
\end{minipage}
 \end{figure}

\subsection{Bounds for $\tilde{u}$ and $\tilde{d}$}

The left-right non-degeneracy bounds for  
 $\tilde{u}$ and $\tilde{d}$ was studied by use of 
 nuclear parity violation in Ref.\cite{NuclParity}. 
Where they compared coefficients of 
 (quark level) meson-nucleon 
 couplings in the 
 SM with those in the SUSY. 
However, studied parameter region was  
 $\mq^2\ll \mg^2<{\cal O}(G_F^{-1})$,  
 which is already experimentally excluded, 
 so that 
 we investigate 
 the left-right non-degeneracy bound in a wider parameter region, 
 besides, without approximations used in Ref.\cite{NuclParity}. 

\begin{figure}[htbp]
  \def\@captype{table}
  \begin{minipage}[r]{.48\textwidth}
  \includegraphics[width=80mm]{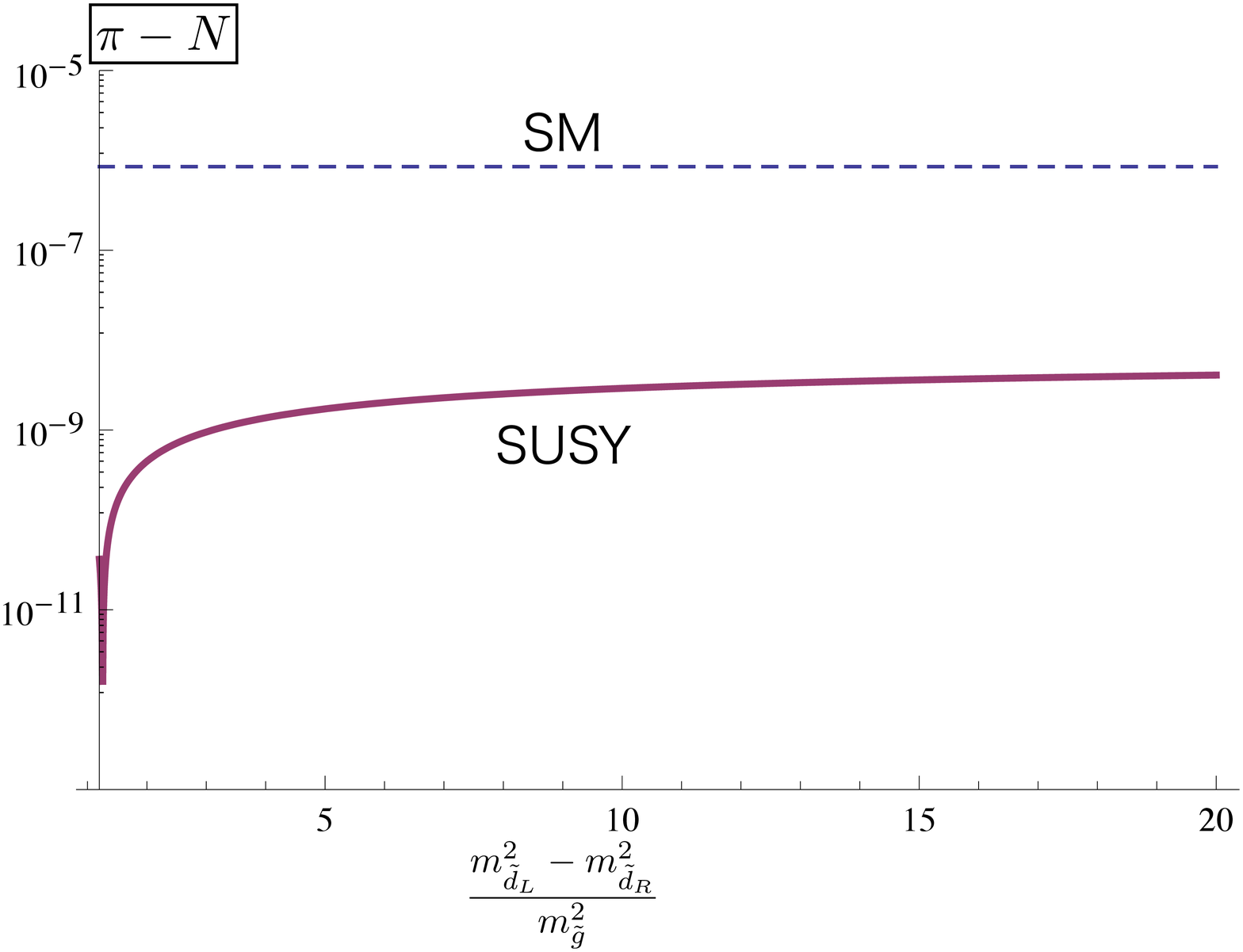}
\caption{\footnotesize $\pi-N$ coupling ($|C^p(\pi)+C^b_-(\pi)|<|C^{\rm SEW}(\pi)|$)
   with $\mg=1000\ {\rm GeV}$, $m_{\tilde{u}_R}=1600\ {\rm GeV}$,
   $m_{\tilde{d}_R}=1700\ {\rm GeV}$, and $m_{\tilde{u}_L}=2000\ {\rm
   GeV}$. The magnitude closes in 
 $7.1\times 10^{-9}$ as 
 $(m_{\tilde{d}_L}^2-m_{\tilde{d}_R}^2)/m_{\tilde{g}}^2\to \infty$.}
 \label{piN}
\end{minipage}
  \hfill
  \begin{minipage}[l]{.48\textwidth}
  \includegraphics[width=80mm]{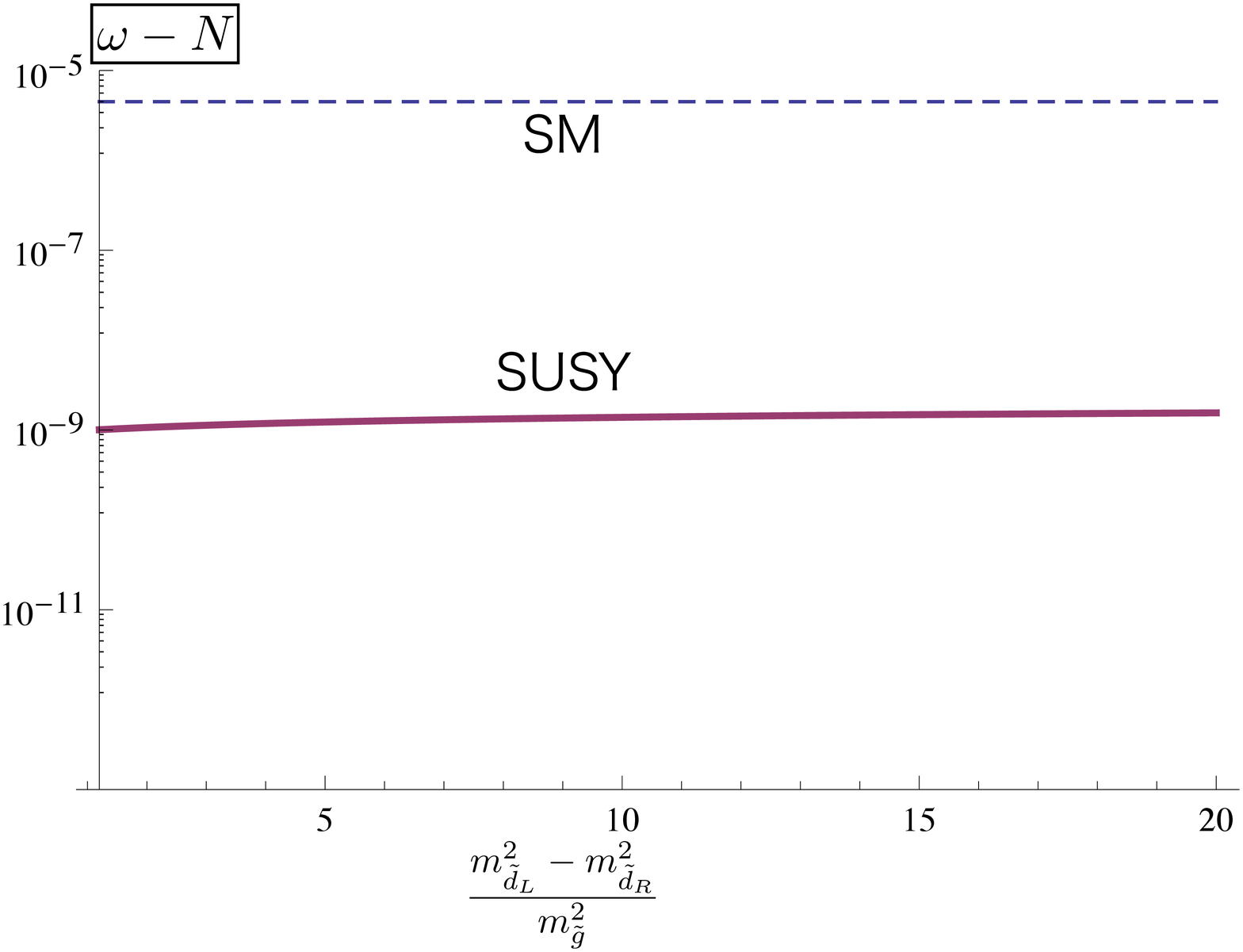}
\caption{\footnotesize $\omega-N$ coupling ($|C^p(\omega)+C^b_+(\omega)|<|C^{\rm
   SEW}(\omega)|$) with $\mg=1000\ {\rm GeV}$, $m_{\tilde{u}_R}=1600\
   {\rm GeV}$, $m_{\tilde{d}_R}=1700\ {\rm GeV}$, and
   $m_{\tilde{u}_L}=2000\ {\rm GeV}$. 
The magnitude closes in 
 $2.1\times 10^{-9}$ as 
 $(m_{\tilde{d}_L}^2-m_{\tilde{d}_R}^2)/m_{\tilde{g}}^2\to \infty$.}
  \label{omegaN}
\end{minipage}
\end{figure}

\begin{figure}[htbp]
  \def\@captype{table}
  \begin{minipage}[c]{.48\textwidth}
  \includegraphics[width=80mm]{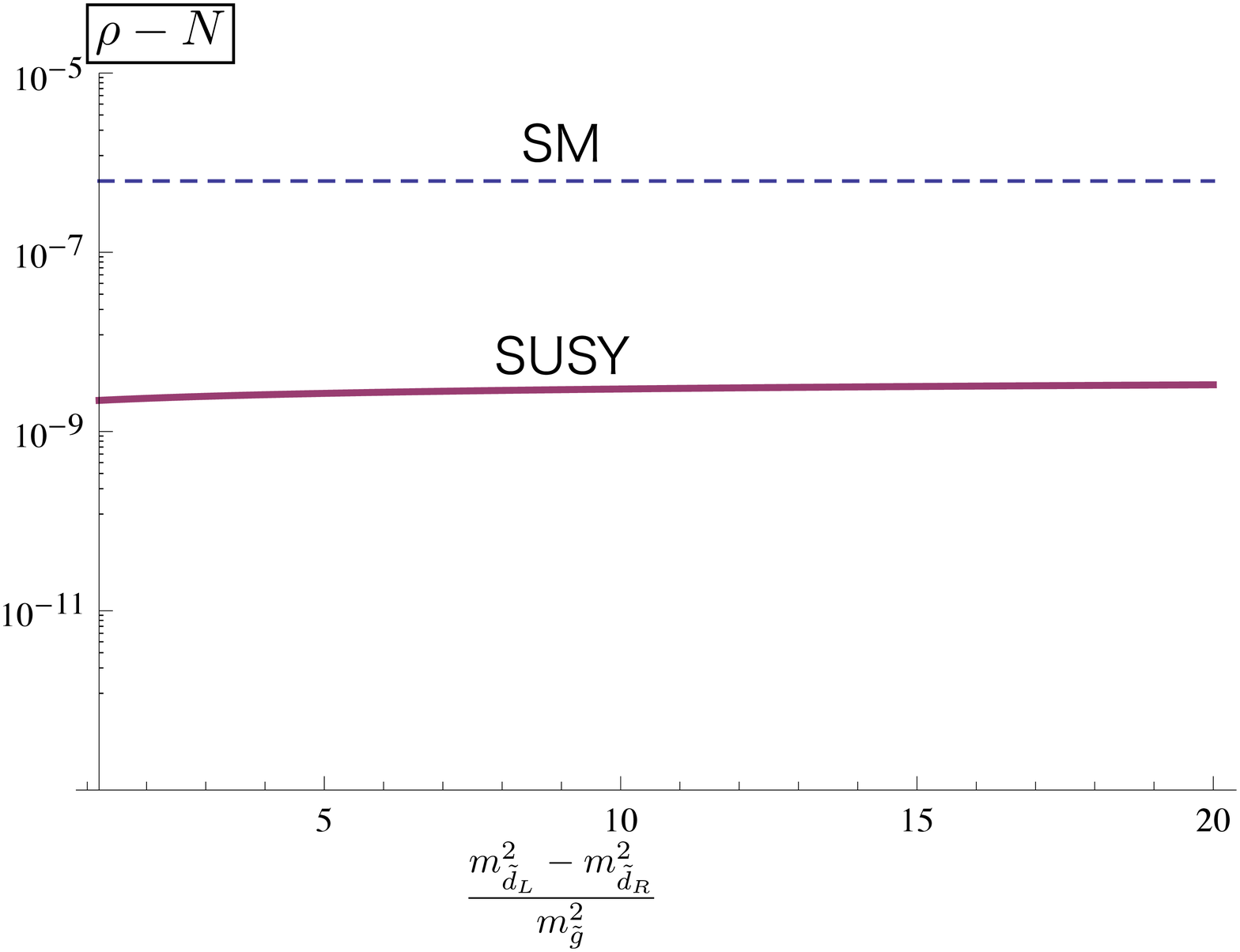}
\caption{\footnotesize $\rho-N$ coupling ($|C^p(\rho)+C^b_+(\rho)|<|C^{\rm
   SEW}(\rho)|$) with $\mg=1000\ {\rm GeV}$, $m_{\tilde{u}_R}=1800\ {\rm
   GeV}$, $m_{\tilde{d}_R}=1700\ {\rm GeV}$, and $m_{\tilde{u}_L}=2000\
   {\rm GeV}$. 
The magnitude closes in 
 $4.4\times 10^{-9}$ as 
 $(m_{\tilde{d}_L}^2-m_{\tilde{d}_R}^2)/m_{\tilde{g}}^2\to \infty$.}  
\label{rhoN}
\end{minipage}
  \hfill
  \begin{minipage}[c]{.48\textwidth}
  \includegraphics[width=80mm]{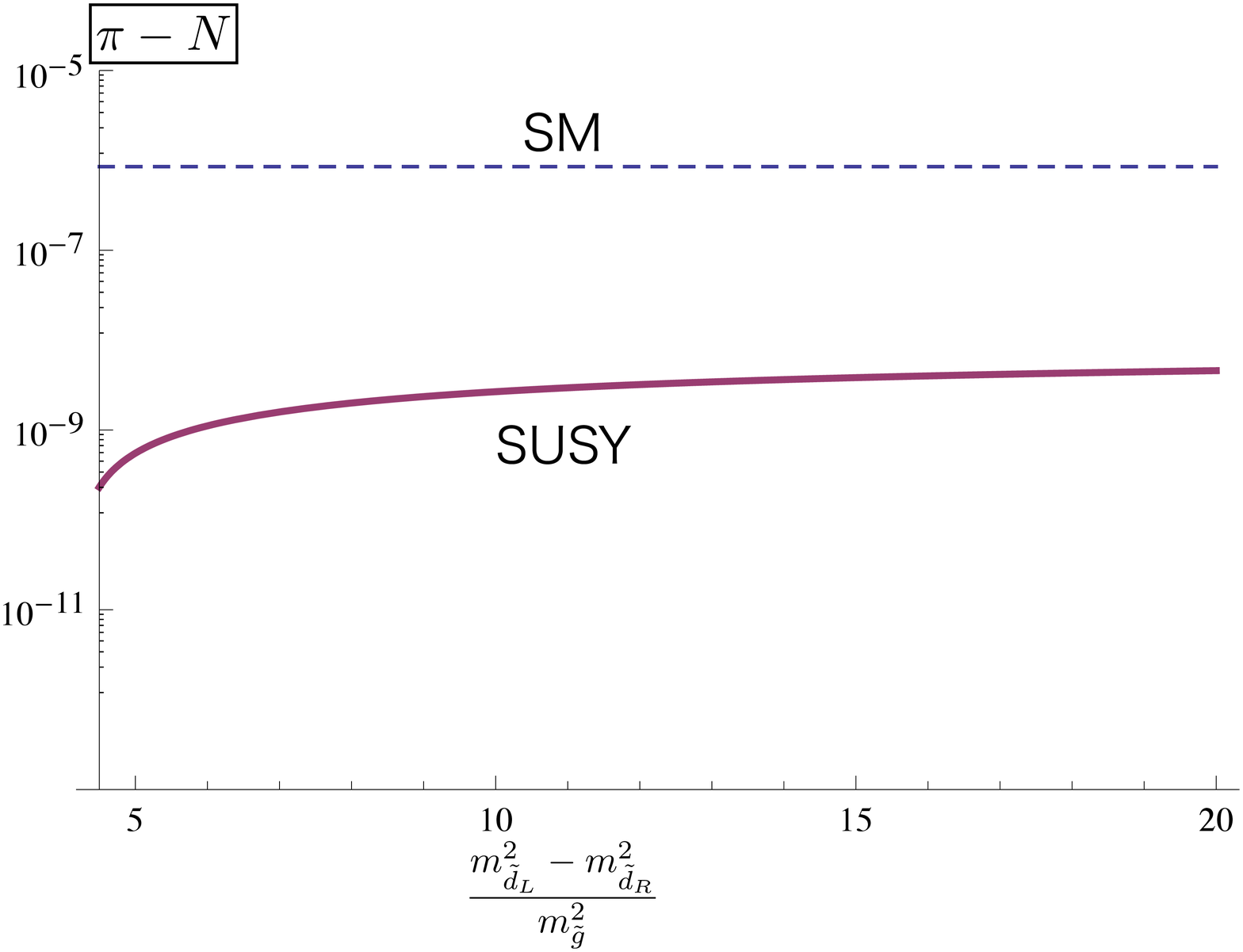}
\caption{\footnotesize $\pi-N$ coupling ($|C^p(\pi)+C^b_-(\pi)|<|C^{\rm SEW}(\pi)|$)
   with $\mg=850\ {\rm GeV}$, $m_{\tilde{u}_R}=860\ {\rm GeV}$,
   $m_{\tilde{d}_R}=870\ {\rm GeV}$, and $m_{\tilde{u}_L}=2000\ {\rm
   GeV}$. 
The magnitude closes in 
 $9.0\times 10^{-9}$ as 
 $(m_{\tilde{d}_L}^2-m_{\tilde{d}_R}^2)/m_{\tilde{g}}^2\to \infty$.}
 \label{piN2}
\end{minipage}
\end{figure}

We use 
 $\pi, \omega,\rho$ and nucleon couplings for  
 the meson-nucleon coupling. 
The notation of our dimension six operators corresponds to 
\begin{align}
\frac{G_2(\mq,m_{\tilde{q}'})}{3\mg^2}=f_1(\mq,m_{\tilde{q}'}),\;\;
\frac{G_1(\mq,m_{\tilde{q}'})}{3\mg^2}=f_2(\mq,m_{\tilde{q}'}), 
\end{align}
 in Ref.\cite{NuclParity}, 
 where we neglect flavor mixings and 
 squark left-right mixings ($A$-terms). 
On the other hand, coefficient of 
 $q$-$q$-$G$ vertex is written by 
\begin{align}
\frac{C(\mq^2/\mg^2)}{\mq^2}
&
=\frac{43 \mg^6-144 \mg^4 \mq^2+153 \mg^2 \mq^4-6 \left(2 \mg^6-9 \mg^2 \mq^4+6 \mq^6\right)
   \log \left(\frac{\mg^2}{\mq^2}\right)-52 \mq^6}{54 \left(\mg^2-\mq^2\right)^4}
\end{align}
in a massless approximation of 
 $u$- and $d$-quarks. 
By using above equations, 
 we calculate bounds from the SM as 
\begin{align}
{\rm (i)}&\;\;\;
|C^p(\pi)+C^b_-(\pi)|<|C^{\rm SEW}(\pi)|,\nonumber\\
{\rm (ii)}&\;\;\;
|C^p(\omega)+C^b_+(\omega)|<|C^{\rm SEW}(\omega)|,\nonumber\\
{\rm (iii)}&\;\;\;
|C^p(\rho)+C^b_+(\rho)|<|C^{\rm SEW}(\rho)|,\nonumber
\end{align}
which are shown in 
 Figs. \ref{piN}, \ref{omegaN}, and \ref{rhoN}. 
$C(\pi)$, $C(\omega)$, and 
 $C(\rho)$
 are parity violating effects (coupling) from 
 $\pi$-, $\omega$-, and $\rho$-nucleon interactions, 
 respectively. 
Indcies $p$ and $b$ stand for 
 penguin and box diagram 
 contributions, respectively. 
Index SEW 
 means an effect from the SM electroweak
 interactions \cite{NuclParity} as 
 $|C^{\rm SEW}(\pi)|=8.5\times 10^{-7},
  |C^{\rm SEW}(\omega)|=4.5\times 10^{-6}$, and 
  $|C^{\rm SEW}(\rho)|=6.2\times 10^{-7}$. 
The factor $c(m_{\tilde{q}})$ is defined by
 $c(m_{\tilde{q}})\equiv C(\mq^2/\mg^2)/\mq^2$, and then  
 \begin{align}
C^p(\pi)
&
=\frac{4}{3}\frac{\alpha_s^2}{12}\rho \left[c(m_{\tilde{u}_R})-c(m_{\tilde{u}_L})-c(m_{\tilde{d}_R})+c(m_{\tilde{d}_L})\right],\\
C^p(\omega)
&
=\frac{1}{3}\frac{\alpha_s^2}{24}\rho \left[c(m_{\tilde{u}_R})-c(m_{\tilde{u}_L})+c(m_{\tilde{d}_R})-c(m_{\tilde{d}_L})\right],\\
C^p(\rho)
&
=\frac{2}{3}\frac{\alpha_s^2}{24}\rho \left[c(m_{\tilde{u}_R})-c(m_{\tilde{u}_L})+c(m_{\tilde{d}_R})-c(m_{\tilde{d}_L})\right],\\
C^b_-(\pi)
&
=-\frac{\alpha_s^2}{27}\rho\left[f_1(m_{\tilde{u}_L},m_{\tilde{d}_R})-f_1(m_{\tilde{u}_R},m_{\tilde{d}_L})-f_2(m_{\tilde{u}_L},m_{\tilde{d}_R})+f_2(m_{\tilde{u}_R},m_{\tilde{d}_L})\right],
\end{align}
\begin{align}
C^b_+(\omega)
&
=-\frac{3\alpha_s^2}{48}\left(\frac{2}{9}+\frac{8}{27}\right)\rho\left[
2f_1(m_{\tilde{u}_L},m_{\tilde{d}_L})-2f_1(m_{\tilde{u}_R},m_{\tilde{d}_R})-f_2(m_{\tilde{u}_L},m_{\tilde{d}_L})+2f_2(m_{\tilde{u}_R},m_{\tilde{d}_R})\right.\nonumber\\
&
-f_1(m_{\tilde{d}_L},m_{\tilde{d}_L})-f_1(m_{\tilde{d}_R},m_{\tilde{d}_R})-f_1(m_{\tilde{u}_L},m_{\tilde{u}_L})-f_1(m_{\tilde{u}_R},m_{\tilde{u}_R})\nonumber\\
&
\left.
+f_2(m_{\tilde{d}_L},m_{\tilde{d}_L})+f_2(m_{\tilde{d}_R},m_{\tilde{d}_R})+f_2(m_{\tilde{u}_L},m_{\tilde{u}_L})+f_2(m_{\tilde{u}_R},m_{\tilde{u}_R})
\right],\\
C^b_+(\rho)
&
=-\frac{\alpha_s^2}{48}\frac{32}{27}
\rho\left[
2f_1(m_{\tilde{u}_L},m_{\tilde{d}_L})-2f_1(m_{\tilde{u}_R},m_{\tilde{d}_R})-f_2(m_{\tilde{u}_L},m_{\tilde{d}_L})+2f_2(m_{\tilde{u}_R},m_{\tilde{d}_R})\right.\nonumber\\
&
-f_1(m_{\tilde{d}_L},m_{\tilde{d}_L})-f_1(m_{\tilde{d}_R},m_{\tilde{d}_R})-f_1(m_{\tilde{u}_L},m_{\tilde{u}_L})-f_1(m_{\tilde{u}_R},m_{\tilde{u}_R})\nonumber\\
&
\left.
+f_2(m_{\tilde{d}_L},m_{\tilde{d}_L})+f_2(m_{\tilde{d}_R},m_{\tilde{d}_R})+f_2(m_{\tilde{u}_L},m_{\tilde{u}_L})+f_2(m_{\tilde{u}_R},m_{\tilde{u}_R})
\right], 
\end{align}
where we take $\rho\sim\sqrt{10}$.

\begin{figure}[htbp]
  \def\@captype{table}
  \begin{minipage}[c]{.48\textwidth}
  \includegraphics[width=80mm]{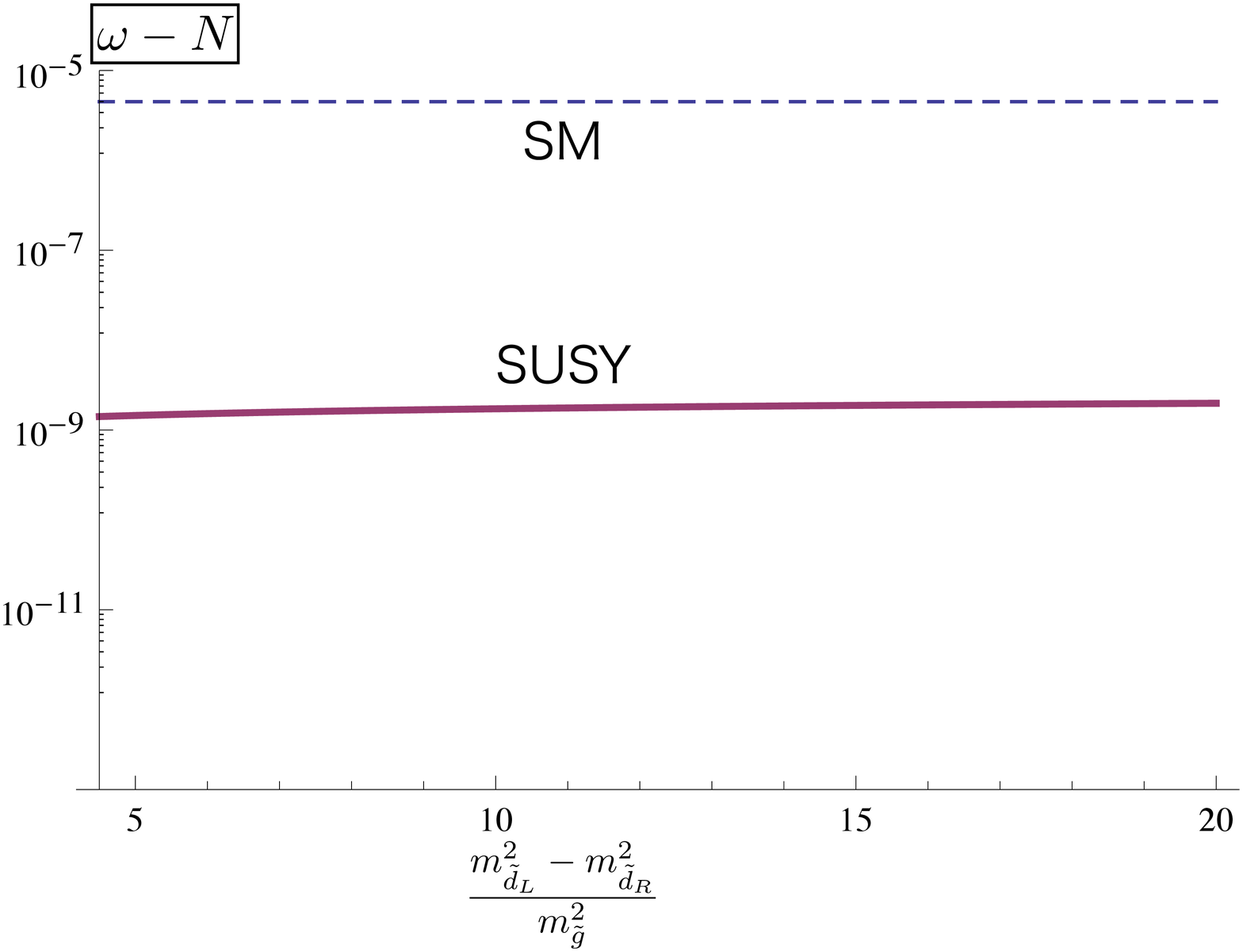}
   \caption{\footnotesize $\omega-N$ coupling ($|C^p(\omega)+C^b_+(\omega)|<|C^{\rm
   SEW}(\omega)|$) with $\mg=850\ {\rm GeV}$, $m_{\tilde{u}_R}=860\ {\rm
   GeV}$, $m_{\tilde{d}_R}=870\ {\rm GeV}$, and $m_{\tilde{u}_L}=2000\
   {\rm GeV}$.
The magnitude closes in 
 $2.7\times 10^{-9}$ as 
 $(m_{\tilde{d}_L}^2-m_{\tilde{d}_R}^2)/m_{\tilde{g}}^2\to \infty$.}
  \label{omegaN2}
\end{minipage}
\hfill
  \begin{minipage}[c]{.48\textwidth}
  \includegraphics[width=80mm]{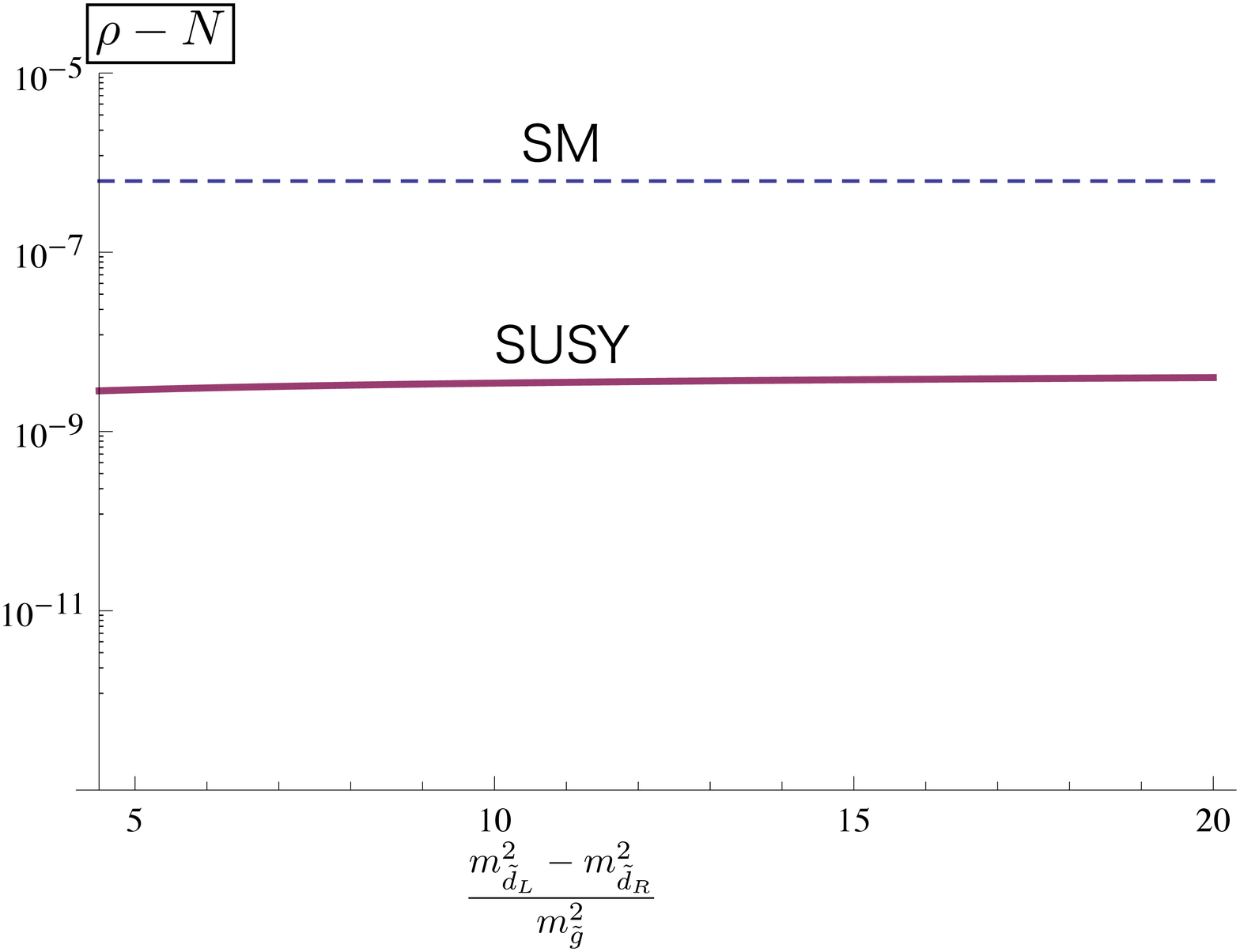}
\caption{\footnotesize $\rho-N$ coupling ($|C^p(\rho)+C^b_+(\rho)|<|C^{\rm
   SEW}(\rho)|$) with $\mg=850\ {\rm GeV}$, $m_{\tilde{u}_R}=860\ {\rm
   GeV}$, $m_{\tilde{d}_R}=870\ {\rm GeV}$, and $m_{\tilde{u}_L}=2000\
   {\rm GeV}$. 
The magnitude closes in 
 $5.6\times 10^{-9}$ as 
 $(m_{\tilde{d}_L}^2-m_{\tilde{d}_R}^2)/m_{\tilde{g}}^2\to \infty$.} 
\label{rhoN2}
\end{minipage}
\end{figure}

In Figs. \ref{piN}, \ref{omegaN}, and \ref{rhoN}, 
 we take sample points  
 which are not excluded by 
 ATLAS experiment\cite{daCosta:2011qk, Aad:2012hm}. 
Under $\mg=1000$ GeV, 
 $m_{\tilde{u}_R}=1600$ GeV, 
 $m_{\tilde{u}_L}=2000$ GeV, 
 and 
 $m_{\tilde{d}_R}=1700$ GeV, 
 we change 
 a value of $(m^2_{\tilde{d}_L}-m^2_{\tilde{d}_R})/\mg^2$ from 
 1.2 for the consistent with the experimental data. 
Unfortunately, 
 in this parameter space,  
 $\tilde{u}$ and $\tilde{d}$ are too heavy to 
 obtain bounds for degeneracies between 
 $m_{\tilde{u}_L}$ and $m_{\tilde{u}_R}$, or, 
 $m_{\tilde{d}_L}$ and $m_{\tilde{d}_R}$. 
On the other hand, 
 when gluino and squarks degenerate within 30 GeV,
 $\pi$-, $\omega$-, and $\rho$-nucleon couplings are
 shown in \ref{piN2}, \ref{omegaN2}, and \ref{rhoN2},
 respectively. 
The magnitude 
 of $(m_{\tilde{d}_L}^2-m_{\tilde{d}_R}^2)/m_{\tilde{g}}^2$ 
 is varied from 4.5 for the consistency 
 with the LHC data. 
In this parameter space,  
 $\tilde{u}$ and $\tilde{d}$ are again too heavy to 
 obtain the bounds. 
The branching ratio is small because 
 SUSY effects always have a loop factor,  
 and it is the reason why 
 there are the asymptotic values in 
 Figs. 7$\sim$12. 

\subsection{Bounds for other sfermions}

Let us comment on
 the bounds for left-right non-degeneracies of 
 other sfermions. 
As for $\tilde{b}$, 
 a total decay width of each bound state of
 $b\ \bar{b}$-meson
 has not experimentally measured yet. 
If we can know the width, 
 the $b\ \bar{b}$-meson system can be
 analyzed, and 
 a bound for a non-degeneracy between 
 $m_{\tilde{b}_L}$ and $m_{\tilde{b}_R}$  
 can be calculated just as the bound between 
 $m_{\tilde{c}_L}$ and $m_{\tilde{c}_R}$  
 was calculated from the charmonium. 
We will calculate the bounds by just replacing 
 $\eta_c \to \eta_b$ ($\eta_b$: $0^{-+}$)  
 and $\chi_{c0}\to\chi_{b0}$ ($\chi_{b0}$: $0^{++}$). 
We hope our method is useful to give a bound 
 between $m_{\tilde{b}_L}$ and $m_{\tilde{b}_R}$ from   
 a future experiments of $B$-physics.  

As for $\tilde{s}$,
 it is difficult to estimate the bound from 
 the same method 
 in section 6.1. 
The reason is as follows. 
If we include 
 a mixing between $d$- and $s$-quarks through 
 the 
 Cabbibo angle, 
 this effect is too small to
 induce the bound between 
 $m_{\tilde{s}_L}$ and $m_{\tilde{s}_R}$  
 because Figs. \ref{piN}, \ref{omegaN}, and 
 \ref{rhoN} can not 
 give bounds for $\tilde{u},\tilde{d}$, too. 
%
On the other hand, 
 if we take $s$-quark as a heavy quark and 
 calculate a quarkonium in NRQCD as $c$-quark, 
 we might have bounds of $\tilde{s}$ 
 for left-right non-degeneracy from 
 parity violating decay mode of
 $\eta(548)$. 
Here,
 $\eta(548)$ is $0^{-+}$ which might have 
 a mixing with $f_0(600)$ ($0^{++}$),  
 if parity violation exists. 
The decay mode of 
 $f_0(600)$ is dominated by $2\pi$. 
Thus, the parity violation induces 
 $\eta(548)\to\pi\pi$, whose experimental bounds are given by 
\begin{align}
{\rm Br}(\eta\to\pi^+\pi^-) < 1.3\times 10^{-5},\;\;\;
{\rm Br}(\eta\to2\pi^0) < 3.5\times 10^{-4}. 
\end{align}
However, these state are 
 not composed only by $s$-quarks but also $u$-, $d$-quarks, so that 
 a valid estimation is
 difficult. 
Also we should remind that 
 mass of $s$-quark is about ten times smaller than 
 that of $c$-quark which is too light to be treated 
 in the NRQCD.

%

Finally, we comment on $sleptons$.  
Lepton flavor violation (LFV) experiments 
 require stringent bounds of non-degeneracy 
 among slepton flavors (generations).
However, the LFV is suppressed when
 slepton masses are 
 heavy enough even if their left- and right-handed slepton masses
 are not degenerate. 
That is, the left-right degeneracy is not required 
 when sleptons are heavy enough. 
This situation is the same for squark sector as above  
 (and also shown in  
 $K^0-\bar{K}^0$ system, 
 where left-right degeneracy is not required 
 with enough heavy squarks).

\section{Summary}

The SUSY SM 
 undergoes parity violation in QCD through 
 chiral    
 quark-squark-gluino 
 interactions with 
 non-degenerate masses between 
 left-handed and right-handed 
 squarks. 
Since experiments have not shown any parity violation  
 in QCD yet,  
 a bound for the mass degeneracy between 
 left-handed and right-handed squarks   
 should exist. 
We have tried to obtain this 
 bound for each squark. 
First of all, we investigated the non-degeneracy bound between 
 $m_{\tilde{c}_L}$ and  $m_{\tilde{c}_R}$ from 
 experimental data of charmonium decay 
 by use of NRQCD.  
Second of all, we estimated the non-degeneracy bounds for
 $\tilde{u}$ and $\tilde{d}$ from nucleon-meson 
 scattering data, 
 and commented on other squarks. 
Unfortunately, 
 our results are below 
  current experimental data, 
 and can not obtain the left-right degeneracy 
 bounds for squark masses. 
We hope our method is useful for obtaining bounds 
 from future experimental data.


\vspace{1cm}

{\large \bf Acknowledegements}\\

\noindent
We thank K. Hanagaki, K. Hikasa, T. Sato, M. Wakamatsu, 
 K. Hagiwara, C.S. Lim, 
 Y. Kuno, T. Yamanaka, and S. Kishimoto,
 for useful and helpful discussions.
 We are also grateful to W. Naylor and C. Kevin for careful reading of the manuscript.
This work is partially supported by Scientific Grant by Ministry of 
 Education and Science, Nos. 20540272, 22011005, 20244028, and 21244036.  

\appendix

\section{Two-body state effective action}

Here we derive the effective action of heavy $q\bar q$-system
in NRQCD, Eq.(\ref{Seff}). 
At leading order in perturbation theory,
we can write an effective QCD Lagrangian as
\begin{align}
S&=
\int_x
\left[
\ol{q} (i\Slash{\partial}-m)q
\right]
+(-i)\int_x\int_y
j^\dagger_\mu(x){\cal D}^{\mu\nu}(x-y)j_\nu(y),\label{QCD}
\end{align}
 where  ${\cal D}^{\mu\nu}(x-y)$ is gluon propagator.
In non-relativistic limit,  
 the gluon propagator 
 induces 
 a (gluon) potential as 
 \begin{align}
 {\cal D}^{\mu\nu}(x-y)&=
 \int\frac{d^4p}{(2\pi)^4}
 \frac{-ig_s^2g^{\mu\nu}}{p^2_0-|\vec{p}|^2}e^{-ip\cdot (x-y)}\nonumber\\
&\simeq
\delta(x^0-y^0)\frac{ig_s^2g^{00}}{4\pi|\vec{x}-\vec{y}| }
\equiv i\delta(x^0-y^0)\frac{V(r)}{C_F}
,
 \end{align}
  where $r=|\vec{x}-\vec{y}|$ and  
 $V(r)\equiv C_Fg_s^2/(4\pi r).$
This is the "Coulomb" potential when energy level of $q\bar q$-system is low
(for example, S-state in $c\bar c$-system, $\eta_c$).
For high energy levels (for example, P-state in $c\bar c$-system, $\chi_c$)
the potential of the heavy $q\bar q$-system $V(r)$ should be well approximated by 
phenomenological potential such a "Coulomb" plus linear as Eq.(\ref{cornell}).
This is because, at longer distance, higher-order perturbation such as 
gluon self interaction gets more important.
In fact,
Refs.\cite{Necco:2001gh, Recksiegel:2001xq, Recksiegel:2002um}
show that the perturbatively calculated QCD potential
agrees with lattice calculations or phenomenologically suggested potential.
 When $q$ is a heavy quark, 
 it is 
 expanded by 
 its mass as 
 \begin{align}
q(x)
&
=\left(
\begin{array}{c}
\varphi e^{-imt}+i\frac{\vec{\nabla}\cdot\vec{\sigma}}{2m}\chi e^{imt}\\
\chi e^{imt}-i\frac{\vec{\nabla}\cdot\vec{\sigma}}{2m}\varphi e^{-imt}
\end{array}
\right).
\label{FWT}
\end{align}
$\varphi$ and $\chi$ denote particle and anti-particle components,
 respectively, and 
 this expansion is so-called 
 Foldy-Wouthuysen-Tani transformation\cite{Foldy:1949wa,Tani1951}.     
Taking 
 a color singlet part in the second term of Eq.(\ref{QCD})  
 (color octet part is the next leading 
 order \cite{Bodwin:1994jh}), 
 we can obtain a NRQCD action, 
\begin{align}
&S_{\rm NRQCD}=\int_x
\left[
\varphi^\dagger\left(i\partial_0+\frac{\nabla^2}{2m}\right)\varphi
+\chi^\dagger\left(i\partial_0-\frac{\nabla^2}{2m}\right)\chi
\right]\nonumber\\
&+\frac{1}{2N_C}\int_x\int_y\delta(x^0-y^0)\left[
\varphi^\dagger(x)\chi(y)V(r)\chi^\dagger(y)\varphi(x)+
\varphi^\dagger(x)\sigma^i\chi(y)V(r)\chi^\dagger(y)\sigma_i\varphi(x)
\right],
\label{nrqcd1}
\end{align}
where 
color factor comes from 
$T^A_{ij}T^A_{kl}
=\frac{C_F}{2N_C}\delta_{il}\delta_{kj}-\frac{1}{N_C}T^A_{il}T^A_{kj} 
$ 
through Fierz transformation. 
Hereafter, we note $N_C$ as a color number,
 which is, of cause, $N_C=3$.  
Next, by inserting the following identities,  
\begin{align}
1&=
\int\prod_{\mu,\nu} {\cal D}s^\mu {\cal D}\phi^{\nu\dagger} \,\,\exp i\int_x\int_y \phi_\mu^{\dagger}(x,y)(s^\mu(x,y)-\varphi^\dagger(x)\sigma^\mu\chi(y)),\\
1&=
\int \prod_{\mu,\nu} {\cal D}s^{\mu\dagger} {\cal D}\phi^{\nu} \,\,\exp i\int_x\int_y
 \phi_\mu(x,y)(s^{\mu\dagger}(x,y)-\chi^\dagger(x)\sigma^\mu\varphi(y)), 
\end{align}
into Eq.(\ref{nrqcd1}),   
 the 
 QCD action becomes 
\begin{align}
S_{\rm NRQCD}&=
\int_x\int_y\left[
\varphi^\dagger(x) K_{\varphi\varphi}\varphi(y)+\chi^\dagger(x) K_{\chi\chi}\chi(y)
-\phi^{\mu\dagger}(x,y)\varphi^\dagger(x)\sigma_\mu\chi(y)-\chi^\dagger(x)\sigma_\mu\varphi(y)\phi^\mu(x,y)\right.\nonumber\\
&\left.+\frac{1}{2N_C}
\delta(x^0-y^0)s^{\mu\dagger}(x,y)V(r)s_\mu(x,y)
+\phi^{\mu\dagger}(x,y)s_\mu(x,y)+s_\mu^\dagger(x,y)\phi^\mu(x,y)
\right], 
\label{nnrr}
\end{align}
where the kinetic terms denote
\begin{align}
\varphi^\dagger(x)
\delta^4(x-y)\left(i\partial_0+\frac{\nabla^2}{2m}\right)
\varphi(y)
&\equiv \varphi^\dagger(x) K_{\varphi\varphi}\varphi(y),\\
\chi^\dagger(x)
\delta^4(x-y)\left(i\partial_0-\frac{\nabla^2}{2m}\right)
\chi(y)
&\equiv \chi^\dagger(x) K_{\chi\chi}\chi(y).
\end{align}

An effective action 
 of the bilocal auxiliary field $\phi^\mu(x,y)$ 
 will be obtained  
 by integrating out $s^\mu, \varphi$, and $\chi$.  
A potential term is induced
 by integrating out $s^\mu$ 
 as 
\begin{align}
-\int_x\int_y\phi^{\mu\dagger}(x,y)\left[
2N_C\delta(x^0-y^0)V^{-1}(r)g_{\mu\nu}
\right]\phi^\nu(x,y).
\end{align}
On the other hand, 
 $\varphi$- and $\chi$-integrations 
 will derive a kinetic term of $\phi^\mu$   
 as shown below. 
We can rewrite the first four terms in Eq.(\ref{nnrr}) as
\begin{align}
 &\int_x\int_y
\left(
\begin{array}{c}
\varphi(x)\\
\chi(x)
\end{array}
\right)^\dagger
\left(
\begin{array}{cc}
K_{\varphi\varphi}&K_{\varphi\chi}\\
K_{\chi\varphi}&K_{\chi\chi}
\end{array}
\right)
\left(
\begin{array}{c}
\varphi(y)\\
\chi(y)
\end{array}
\right),
\end{align}
where $K_{\varphi\chi}$ and $K_{\varphi\chi}$
 are denoted as 
$K_{\varphi\chi}=-\phi^{\mu\dagger}(x,y)\sigma_\mu$ and 
$K_{\chi\varphi}=-\sigma_\mu\phi^\mu(x,y)$.
Then,
by integrating out 
 $\varphi$ and $\chi$ in Eq.(\ref{nnrr}), 
 we can obtain the term 
 \begin{align}
i{\rm Tr} \log
\left(
\begin{array}{cc}
K_{\varphi\varphi}&K_{\varphi\chi}\\
K_{\chi\varphi}&K_{\chi\chi}
\end{array}
\right)
&\simeq
i{\rm Tr}\log
\left(
\begin{array}{cc}
K_{\varphi\varphi}&0\\
0&K_{\chi\chi}
\end{array}
\right)
+
i\sum_{n=1}^\infty \frac{(-1)^{n-1}}{n}
{\rm Tr}\left(
\begin{array}{cc}
0&K_{\varphi\varphi}^{-1}K_{\varphi\chi}\\
K_{\chi\chi}^{-1}K_{\chi\varphi}&0
\end{array}
\right)^n\label{Tr-log}.
\end{align}
The ${\rm Tr}\log$ is expanded  in Eq.(\ref{Tr-log}), 
where 
 $n=1$ is vanished by a trace,
 and the leading term is coming from $n=2$. 
After taking traces of
 spinor, 
 color, and coordinate indices, 
 the leading term 
 in Eq.(\ref{Tr-log}) becomes
\begin{align}
&\frac{-i}{2}{\rm Tr}
\left(
\begin{array}{cc}
K_{\varphi\varphi}^{-1}K_{\varphi\chi}K_{\chi\chi}^{-1}K_{\chi\varphi}&0\\
0&K_{\chi\chi}^{-1}K_{\chi\varphi}K_{\varphi\varphi}^{-1}K_{\varphi\chi}
\end{array}
\right)\nonumber\\
&=
-iN_C\int_x\int_y\int_z\int_w
{\rm Tr}_{\rm spin}
 K_{\varphi\varphi}^{-1}(x,y)K_{\varphi\chi}(y,z)K_{\chi\chi}^{-1}(z,w)K_{\chi\varphi}(w,x), 
 \label{tr}
\end{align}
where propagators are given by 
\begin{align}
K_{\varphi\varphi}^{-1}(x,y)&=
\int\frac{d^4p}{(2\pi)^4} 
\frac{1}{p^0-\frac{\vec{p}^2}{2m}+i\epsilon}
e^{-ip(x-y)}\delta^{\alpha\beta},\nonumber\\
K_{\chi\chi}^{-1}(x,y)&=
-\int\frac{d^4q}{(2\pi)^4}
 \frac{1}{q^0-\frac{\vec{q}^2}{2m}+i\epsilon}e^{iq(x-y)}\delta^{\alpha\beta}.
\end{align}
We use a center of mass coordinate 
 $X^\mu$ and relative
 coordinate $(0, \vec{r})^\mu$ 
 as 
 $x^\mu=X^\mu+\frac{1}{2}(0,\vec{r})^\mu$ and 
 $y^\mu=X^\mu-\frac{1}{2}(0,\vec{r})^\mu$.  
The relative coordinate does not have $time$-component, 
 since $\phi^\mu(x,y)$ is a coincident 
 bilocal field 
 for $x$ and $y$.  
Then,
 $\phi^\mu(x,y)$ 
 is represented by 
\begin{align}
 \phi^\mu(x,y)\equiv \phi^\mu_X(\vec{r})
=\int_k\int_l\phi^\mu_k(\vec{l})e^{-ikX}e^{-il_\mu(0,\vec{r})^\mu}
=\int_k\int_{l^0}\int_{\vec{l}}\phi^\mu_k(\vec{l})e^{-ikX}e^{i\vec{l}\cdot\vec{r}},
\end{align}
with their momentums as 
$p^\mu=(\frac{k^0}{2}+l^0, \frac{\vec{k}}{2}+\vec{l})$ and
$q^\mu=(\frac{k^0}{2}-l^0, \frac{\vec{k}}{2}-\vec{l})$. 
In this frame, 
 Eq.(\ref{tr}) is written as 
\begin{align}
&-2N_C\int_k\int_l\int_{\vec{r}}\int_{\vec{s}}
\frac{1}{\left[\frac{k^0}{2}+l^0-\frac{(\vec{k}/2+\vec{l})^2}{2m}+i\epsilon \right]\left[\frac{k^0}{2}-l^0-\frac{(\vec{k}/2-\vec{l})^2}{2m}+i\epsilon \right]}
\phi^{\mu\dagger}_k(\vec{r})\phi_{\mu k}(\vec{s})e^{-i\vec{l}\cdot(\vec{r}+\vec{s})}
,\nonumber\\
\end{align}
and we obtain
\begin{align}
&-2iN_C\int_k\int_{\vec{l}}\int_{\vec{r}}\int_{\vec{s}}
\frac{1}{k^0-\frac{\vec{k}^2}{4m}-\frac{\vec{l}^2}{m}}\phi^{\mu\dagger}_k(\vec{r})\phi_{\mu k}(\vec{s})
e^{-i\vec{l}\cdot(\vec{r}+\vec{s})}
\end{align}
by integrating $l^0$.
Then,
 the effective action of $\phi^\mu$ 
 is given by 
\begin{align}
S_{\rm eff}&=
\int_X \int_{\vec{r}}
\phi^{\mu\dagger}_X(\vec{r})
\left[
\frac{1}{V(r)}-\frac{1}{K_X(r)}
\right]
\phi_{\mu X}(\vec{r}),
\label{438}
\end{align}
where $K_X(r)\equiv i\partial^0_X-\frac{\nabla^2_X}{4m}-\frac{\nabla^2_r}{m}$.
We omit overall factor $2N_C$ by use of 
 normalization of the field. 
Note that a Green function
 $\langle \phi^\mu_X(\vec{r})\phi^{\nu\dagger}_Y(\vec{s})\rangle$ 
 is given by 
\begin{align}
\langle \phi^\mu_X(\vec{r})\phi^{\nu\dagger}_Y(\vec{s})\rangle
&
\equiv\left[
V^{-1}-K^{-1}
\right]^{-1}_{\mu\nu}(X,\vec{r};Y,\vec{s})\nonumber\\
&
=V(r)g_{\mu\nu}\delta^4(X-Y)\delta^3(\vec{r}-\vec{s})
+\left[
V(K-V)^{-1}V
\right]_{\mu\nu}(X,\vec{r};Y,\vec{s}). 
\label{Green1}
\end{align}
In asymptotic states, $X\neq Y$, 
 the first term vanishes. 
The second term is what we want, and 
 $V$ is rotated out by field redefinition, 
 then 
 Eq.(\ref{Green1}) becomes 
\begin{align}
\langle \phi^\mu_X(\vec{r})\phi^{\nu\dagger}_Y(\vec{s})\rangle
&
=
\left[
(K-V)^{-1}
\right]_{\mu\nu}(X,\vec{r};Y,\vec{s}).
\end{align}
This means 
 that 
 the 
 effective action in Eq.(\ref{438}) 
 can be rewritten as 
\begin{align}
S_{\rm eff}&=
\int_X \int_{\vec{r}}
\phi^{\mu\dagger}_X(\vec{r})
\left[
K_X(r)-V(r)
\right]
\phi_{\mu X}(\vec{r}). 
\label{NRQCD-Seff}
\end{align}
This is the effective action of Eq.(\ref{Seff}).
We should notice that 
 this form is correct when 
 asymptotic states exist\cite{Politzer:1980me}    
 and $\phi_X^\mu(\vec{r})$ is an
 on-shell state. 
 

\section{Dimension six operators from SUSY}


We calculate dimension six operators 
 by intgrating out sparticles 
 in the framework $R$-parity conservation.  
By integrating out SUSY particles,
 we can obtain higher order gauge invariant
 operators in terms of the SM fields.
We calculate them up to $\mathcal{O}(\alpha_s^2)$ 
 and neglect $\mathcal{O}(\alpha_s \alpha_y)$, 
 where $\alpha_s$ ($\alpha_y$)
 is $g_s^2/4\pi$ ($y^2/4\pi$)
 with 
 a QCD (Yukawa) coupling, $g_s$ ($y$). 
It is because 
 Yukawa couplings, $y$s, which we deal with are all small\footnote{
We do not analyze $\tilde{t}$ in this paper.},  
 and 
 up to this order, 
 sfermion left-right mixings are negligible. 
%
In the NRQCD, 
 there are three types of dimension six operators,
 ${\cal O}^{(1)}_{4F}, {\cal O}^{(8)}_{4F}$, and ${\cal O}_{qqG}$,
 which can contribute the parity violation in QCD.
Here ${\cal O}^{(1)}_{4F}$
 and ${\cal O}^{(8)}_{4F}$ are 
 color singlet and octet 4-Fermi operator, respectively. 
Other dimension six 
 operators such as
 $q$-$q$-$G$-$G$ and $q$-$q$-$G$-$G$-$G$ vertexes 
 are next leading order in the NRQCD, 
 so we neglect them in the following discussions. 

For ${\cal O}^{(1)}_{4F}$ and ${\cal O}^{(8)}_{4F}$, 
 they 
 are
 given by \cite{Haba:2011vi} 
\begin{align}
{\cal O}^{(1)}_{4F}&=\frac{12g_s^4}{192\pi^2}\sum_{q,q'}^{\rm flavor}
\left[
C_{LL}
\left(
\bar{q}\gamma^\mu P_Lq
\right)
\left(
\bar{q}'\gamma_\mu P_Lq'
\right)
+
C_{RR}
\left(
\bar{q}\gamma^\mu P_Rq
\right)
\left(
\bar{q}'\gamma_\mu P_Rq'
\right)\right.
\nonumber\\
&\qquad\qquad\left.+
C_{LR}
\left(
\bar{q}\gamma^\mu P_Lq
\right)
\left(
\bar{q}'\gamma_\mu P_Rq'
\right)
+
C_{RL}
\left(
\bar{q}\gamma^\mu P_Rq
\right)
\left(
\bar{q}'\gamma_\mu P_Lq'
\right)
\right],\label{SUSY-4F-1}
\\
{\cal O}^{(8)}_{4F}&=\frac{12g_s^4}{192\pi^2}\sum_{q,q'}^{\rm flavor}
\left[
D_{LL}
\left(
\bar{q}T^a\gamma^\mu P_Lq
\right)
\left(
\bar{q}'T^a\gamma_\mu P_Lq'
\right)
+
D_{RR}
\left(
\bar{q}T^a\gamma^\mu P_Rq
\right)
\left(
\bar{q}'T^a\gamma_\mu P_Rq'
\right)\right.
\nonumber\\
&\qquad\qquad\left.+
D_{LR}
\left(
\bar{q}T^a\gamma^\mu P_Lq
\right)
\left(
\bar{q}'T^a\gamma_\mu P_Rq'
\right)
+
D_{RL}
\left(
\bar{q}T^a\gamma^\mu P_Rq
\right)
\left(
\bar{q}'T^a\gamma_\mu P_Lq'
\right)
\right],\label{SUSY-4F-8}
\end{align}
where coefficients are 
 \begin{align}
C_{LL}&=\frac{2}{9}[f_1(m_{\tilde{q}_L},m_{\tilde{q}'_L})+f_2(m_{\tilde{q}_L},m_{\tilde{q}'_L})], \label{22}\\
C_{RR}&=\frac{2}{9}[f_1(m_{\tilde{q}_R},m_{\tilde{q}'_R})+f_2(m_{\tilde{q}_R},m_{\tilde{q}'_R})],\\
C_{LR}&=-\frac{2}{9}[f_1(m_{\tilde{q}_R},m_{\tilde{q}'_L})-f_2(m_{\tilde{q}_L},m_{\tilde{q}'_R})],\\
C_{RL}&=-\frac{2}{9}[f_1(m_{\tilde{q}_L},m_{\tilde{q}'_R})+f_2(m_{\tilde{q}_L},m_{\tilde{q}'_R})], \label{24}
\end{align}
\begin{align}
D_{LL}&=-\frac{1}{3}f_1(m_{\tilde{q}_L},m_{\tilde{q}'_L})-\frac{7}{6}f_2(m_{\tilde{q}_L},m_{\tilde{q}'_L}),\\
D_{RR}&=-\frac{1}{3}f_1(m_{\tilde{q}_R},m_{\tilde{q}'_R})-\frac{7}{6}f_2(m_{\tilde{q}_R},m_{\tilde{q}'_R}),\\
D_{LR}&=-\frac{7}{6}f_1(m_{\tilde{q}_L},m_{\tilde{q}'_R})-\frac{1}{3}f_2(m_{\tilde{q}_L},m_{\tilde{q}'_R}),\\
D_{RL}&=-\frac{7}{6}f_2(m_{\tilde{q}_R},m_{\tilde{q}'_L})-\frac{1}{3}f_1(m_{\tilde{q}_R},m_{\tilde{q}'_L}),
\end{align}
\begin{align}
f_1(\mq,m_{\tilde{q}'})
&
=\int^1_0dy\int^1_0dz\frac{yz^2}{(\mg^2-\mq^2)yz+(\mq^2-m_{\tilde{q}'}^2)z+m_{\tilde{q}'}},\label{Feyn1}\\
f_2(\mq,m_{\tilde{q}'})
&
=\int^1_0dy\int^1_0dz\frac{\mg^2yz^2}{[(\mg^2-\mq^2)yz+(\mq^2-m_{\tilde{q}'}^2)z+m_{\tilde{q}'}]^2}.\label{Feyn2}
 \end{align}
On the other hand, 
 ${\cal O}_{qqG}$ 
 is
 given by 
\begin{align}
{\cal O}_{qqG}
&
=\frac{g_s^3}{96\pi^2}
\int\frac{d^4k_1}{(2\pi)^4}\frac{d^4k_2}{(2\pi)^4}\frac{d^4k_3}{(2\pi)^4}
(2\pi)^4\delta^4(-k_1+k_2+k_3)
\ol{q}(k_2)T^aE^\mu_{L,R} G^a_\mu(k_3)P_{L,R}q(k_1),\label{SUSY-qqG}\\
E^\mu_L
&
\equiv E^\mu(\mq=m_{\tilde{q}_L}),\nonumber\\
&
=\{e_1(m_{\tilde{q}_L})\Slash{k}_1+e_2(m_{\tilde{q}_L})\Slash{k}_2\}k^\mu_1
+\{e_1(m_{\tilde{q}_L})\Slash{k}_2+e_2(m_{\tilde{q}_L})\Slash{k}_1\}k^\mu_2\nonumber\\
&+\{e_3(m_{\tilde{q}_L})(k_1^2+k_2^2)-e_4(m_{\tilde{q}_L})k_1\cdot k_2\}\gamma^\mu
-e_5(m_{\tilde{q}_L})i\epsilon^{\alpha\beta\mu\nu}\gamma_5\gamma_\nu
k_{1\alpha}k_{2\beta},\label{SUSY-qqG}\\
E^\mu_R&=
E^\mu(m_{\tilde{q}_R}),
\end{align}
\begin{align}
&e_1(\mq)\nonumber\\
&=
\frac{107\mg^6-495\mg^4\mq^2+477\mg^2\mq^4-89\mq^6-6(\mg^6+3\mg^4\mq^2-54\mg^2\mq^4+18\mq^6)\log(\mg^2/\mq^2)}{18(\mg^2-\mq^2)^4},\\
&e_2(\mq)\nonumber\\
&=
\frac{-203\mg^6+351\mg^4\mq^2-189\mg^2\mq^4+41\mq^6+6(\mg^6+51\mg^4\mq^2-54\mg^2\mq^4+18\mq^6)\log(\mg^2/\mq^2)}{18(\mg^2-\mq^2)^4},\\
&e_3(\mq)=e_2(\mq),\\
&e_4(\mq)\nonumber\\
&=
\frac{-155\mg^6+423\mg^4\mq^2-333\mg^2\mq^4+65\mq^6+6(\mg^6+27\mg^4\mq^2-54\mg^2\mq^4+18\mq^6)\log(\mg^2/\mq^2)}{9(\mg^2-\mq^2)^4},\\
&e_5(\mq)=\frac{9(\mg^4-\mq^4-2\mg^2\mq^2\log(\mg^2/\mq^2))}{(\mg^2-\mq^2)^3}.
\end{align}
We use these QCD dimension six operators  
 in order to obtain the non-degeneracy bounds 
 of left-right squark masses.


\end{document}